\documentclass[10pt]{article}

\usepackage{amssymb,amsmath}

\newcommand{\zt}[1]{{#1}}
\newcommand{\z}[1]{{#1}}
\newcommand{\Ad}{C}
\newcommand{\RRR}{\mathbb{R}}

\newcommand{\NNN}{\mathbb{N}}
\newcommand{\CCC}{\mathbb{C}}

\newcommand{\bx}{\boldsymbol x}

\newcommand{\bxp}{\boldsymbol x'}
\newcommand{\bkp}{\boldsymbol k'}

\newcommand{\ba}{\boldsymbol a}
\newcommand{\bq}{\boldsymbol q}
\newcommand{\bk}{\boldsymbol k}

\newcommand{\bko}{\boldsymbol k_0}
\newcommand{\by}{\boldsymbol y}
\newcommand{\be}{\boldsymbol e}

\newcommand{\gre}{\frac{e^{ik|\bx-\bxp|}}{|\bx-\bxp|}}

\newcommand{\bbq}{\boldsymbol Q}
\newcommand{\bj}{\boldsymbol j}
\newcommand{\bsig}{\boldsymbol \sigma}
\newcommand{\bv}{\boldsymbol v}

\newcommand{\bn}{\boldsymbol n}
\newcommand{\bom}{\boldsymbol \omega}
\newcommand{\nd}{N_{\text{det}}}

\newcommand{\nsi}{N_{\text{sig}}}

\newcommand{\pin}{\psi_{\text{in}}}
\newcommand{\pout}{\psi_{\text{out}}}
\newcommand{\cout}{\chi_{\text{out}}}
\newcommand{\pouth}{\widehat{\psi}_{\text{out}}}
\newcommand{\couth}{\widehat{\chi}_{\text{out}}}

\newcommand{\ran}{\text{Ran}}
\newcommand{\dom}{\text{D}}

\newcommand{\dteb}{\tau}
\newcommand{\dte}{\tau}

\newcommand{\kfac}{\kappa}

\newcommand{\gprime}{\mathcal{G}^0}
\newcommand{\g}{\mathcal{G}}
\newcommand{\ghut}{\mathcal{G}^{+}}
\newcommand{\ltr}{L^2(\RRR^3)}
\newcommand{\lxr}{\langle x\rangle}
\newcommand{\lkr}{\langle k\rangle}

\newcommand{\supp}{\operatorname{supp}}
\newcommand{\diver}{\operatorname{div}}

\newcommand{\limmean}{\operatorname{l.i.m.}}

\newcommand{\dti}{[0,\infty)}
\newcommand{\siglimdiff}{\sigma^{\text{diff}}}

\newcommand{\siglim}{\sigma}
\newcommand{\sigemp}{\sigma_{\text{emp}}}

\newcommand{\Lambdastar}{\Lambda^{\star}}
\newcommand{\lambdastar}{\lambda^{\star}}

\newcommand{\chara}{\chi_B}
\newcommand{\charac}{\chi}

\newcommand{\GammaL}{Y_L}

\newcommand{\Leps}{L^{\epsilon}}
\newcommand{\Aeps}{A^{\epsilon}}

\newcommand{\Deps}{D^{\epsilon}}
\newcommand{\Reps}{\frac{\Deps}{2}}
\newcommand{\psieps}{\psi^{\epsilon}}
\newcommand{\psiyeps}{\psi_{\by}^{\epsilon}}

\newcommand{\yp}{y_p}
\newcommand{\byp}{\by_p}

\newcommand{\bkpl}{\bk_p}
\newcommand{\schwarz}{\mathcal{S}(\RRR^3)}

\newcommand{\bkplus}{\bk_+}
\newcommand{\con}{c}

\newcommand{\ndstar}{N^{\star}}

\newsavebox{\proofnumber}

\newenvironment{proof}[1]{\vspace{0,5cm}\parindent=0pt\sbox{\proofnumber}{\sl #1}{\sl Proof}\usebox{\proofnumber}.\parindent=15pt }{\vspace{0,5cm}}

\newtheorem{theorem}{Theorem}

\newcommand{\bthe}{\begin{theorem}\hspace{-1.1ex}{\bf .}\hspace{2ex}}
\newcommand{\ethe}{\end{theorem}}

\newcommand{\bpro}{\begin{proof}}

\newcommand{\epro}{$\blacksquare$\end{proof}}

\newtheorem{lemma}{Lemma}

\newcommand{\blem}{\begin{lemma}\hspace{-1.1ex}{\bf .}\hspace{2ex}}
\newcommand{\elem}{\end{lemma}}

\newtheorem{remark}{\sl Remark}

\newcommand{\brem}{\begin{remark}\rm\hspace{-1.1ex}{.}\hspace{2ex}}
\newcommand{\erem}{\end{remark}}

\newtheorem{definition}{Definition}

\newcommand{\bde}{\begin{definition}\hspace{-1.1ex}{\bf .}\hspace{2ex}}
\newcommand{\ede}{\end{definition}}

\newtheorem{corollary}{Corollary}

\newcommand{\bcor}{\begin{corollary}\hspace{-1.1ex}{\bf .}\hspace{2ex}}
\newcommand{\ecor}{\end{corollary}}

\newtheorem{proposition}{Proposition}

\newcommand{\bprop}{\begin{proposition}\hspace{-1.1ex}{\bf .}\hspace{2ex}}
\newcommand{\eprop}{\end{proposition}}

\newcounter{cond}

\newenvironment{cond}[0]{\refstepcounter{cond}
\begin{itemize}\item[\bfseries{A\arabic{cond}.}]}{\end{itemize}}

\newcommand{\bcond}{\begin{cond}}
\newcommand{\econd}{\end{cond}}

\begin{document}
\parindent=0pt

\thispagestyle{empty}


{\huge{\bfseries{A microscopic derivation of
the quantum }}}

{\huge{\bfseries{mechanical formal scattering cross section}}}

\vspace{1cm}

D. D\"urr$^1$, S. Goldstein$^2$, T. Moser$^1$, N.
Zangh{\`{\i}}$^3$

\begin{list}{1}{\setlength{\labelwidth}{2em}\setlength{\leftmargin}{1em}}
\item[1] Mathematisches Institut der Universit\"at M\"unchen,
\\Theresienstr. 39, 80333 M\"unchen, Germany \item[2] Department
of Mathematics, Rutgers University, New Brunswick, NJ 08903, USA
\item[3] Dipartimento di Fisica, Universit{\`a} di Genova, Sezione
INFN Genova,\\Via Dodescanesco 33, 16146 Genova, Italy
\end{list}

March 13, 2006\vspace{1cm}

{\bfseries{\small{Abstract. }}}{\small{We prove that the empirical
distribution of crossings of a ``detector'' surface by scattered
particles converges in appropriate limits to the scattering cross
section computed by stationary scattering theory. Our result,
which is based on Bohmian mechanics and the flux-across-surfaces
theorem, is the first derivation of the cross section starting
from first microscopic principles.}}

\parindent=15pt
\section{Introduction}\label{chintro}

The central quantity in a scattering experiment is the empirical
cross section, which reflects the number of particles that are
scattered in a given solid angle per unit time. In this paper we
shall derive the theoretical prediction for the cross section
starting from a microscopic model describing a realistic
scattering situation. We confine ourselves to the case of
potential scattering of a nonrelativistic, (spinless) quantum
particle and leave the many-particle case for future research.
This paper is in fact a technical elaboration and continuation of
our article ``Scattering theory from microscopic first principles"
\cite{duerr:00}.

The common approaches to the foundations of scattering theory take
for granted that ``an experimentalist generally prepares a \z{state}
\dots\ at $t\to -\infty$, and then measures what this state looks
like at $t\to +\infty$'' (cf. \cite{weinberg:96}, p. 113), meaning
that the asymptotic expressions are ``all there is,'' as if they
are not the asymptotic expressions of some other formula, however
complicated, describing the scattering situation as it really is,
namely happening at finite distances and at finite times. Thus a
truly microscopic derivation starting from first principles must
provide firstly a formula for the empirical cross section, which
by the law of large numbers approximates its expectation value,
and which is computed from the underlying theory. Secondly, that
formula should apply to the realistic \zt{finite-times and} finite-distances situation,
from which eventually the usual Born formula should emerge by
taking appropriate limits.\footnote{For a detailed discussion of
the scattering regime see \cite{duerr4:04}.}

We shall present a Bohmian analysis of the scattering cross
section. With a particle trajectory we can ask for example whether
or not that trajectory eventually crosses a distant spherical
surface and if it does when and where it first crosses that
surface. Similarly, for a beam of particles we can ask for the
number of particles in the beam that first crosses the surface in
a given solid angle $\Sigma.$ From a Bohmian perspective it
appears reasonable to identify this number with detection events
in a scattering experiment. We thus model in this paper the
measured cross section \z{using} the number $\ndstar(\Sigma)$ of first
crossings of $\Sigma.$ This will of course depend on many
parameters encoding the experimental setup, e.g. the distances $R$
and $L$ of the detector and the particle source from the
scattering center, the details of the beam including its profile
$A$ \z{and} the wave functions of the particles in the \z{beam,} as well as on
the length of the time interval $\dte$ during \z{which} the particles are
emitted. We shall show in this paper that when these parameters
are suitably scaled, $\frac{\ndstar(\Sigma)}{\dteb}$ is well
approximated by the usual Born formula for the scattering cross
section in terms of the $T$-matrix, \z{i.e.,}
\begin{align}\label{rums}
\lim\frac{\ndstar(\Sigma)}{\dteb}=&16\pi^4\int\limits_{\Sigma}|T(k_0\bom,\bko)|^2d\Omega,
\end{align}
\z{where $\hbar \bko$ is the initial momentum of the particles.}

The paper is organized as follows: We collect first some
mathematical notions and facts as well as recent results of
scattering theory. In Section \ref{chflux} we define the relevant
random variables associated with the surface-crossings of a single
particle and relate \z{their} distribution to the quantum probability
current density. In Section \ref{chmodel} we model the beam by a
suitable point process and in Section \ref{chdefcross} we define
$\ndstar(\Sigma)$ in terms of this point process. A precise
description of the limit procedure will be presented in Section
\ref{chregime}. Our main results, Theorem \ref{thedasisses} and
\ref{thedasisseslight}, are stated in Section \ref{chlimit} and
are proven in Section \ref{chformula}.

\section{The mathematical framework of potential scattering}\label{chsetup}

\sectionmark{The mathematical framework}
We list those results of scattering theory (e.g. \cite{amrein:77,
duerr:01,duerr1:04, ikebe:60, kato:51, pearson:88, simon3:79,
simon1:80, teufel2:99}) which are essential for the proof of Theorem \ref{thedasisses} and Theorem \ref{thedasisseslight} in Section \ref{chformula}.

We use the usual description of a nonrelativistic spinless
one-particle system by the Hamiltonian $H$\label{hamiltonian} (we
use natural units $\hbar=m=1$),
\begin{equation*}
  H:=-\frac{1}{2}\Delta+V(\bx)=:H_0+V(\bx),
\end{equation*}
with the \zt{real-valued potential}\label{potential}
$V\in (V)_n$, defined as follows:

\bde\label{defpot} $V$ is in $(V)_n$, n=2,3,4,..., if
\begin{itemize}

\item[(i)] $V\in \ltr$,\item[(ii)] $V$ is locally H\"older continuous except, perhaps, at a finite
number of singularities, 
\item[(iii)] there exist positive numbers
$\delta,\;C,\;R_0$ such that
\begin{equation*}
  |V(\bx)|\leq C\lxr^{-n-\delta}\text{ for }x\geq R_0,
\end{equation*}

where $\langle\cdot\rangle:=(1+(\cdot)^2)^{\frac{1}{2}}.$
\end{itemize}

\ede
Under these conditions (see e.g. \cite{kato:51})  $H$ is
self-adjoint on the domain D($H)=$D($H_0)=\{f\in
\ltr:\int|k^2\widehat{f}(\bk)|^2d^3k<\infty\}$ ($k=|\bk|)$, where
$\widehat{f}:=\mathcal{F}f$ is the Fourier transform
\begin{equation}\label{fourier}
  \widehat{f}(\bk):=(2\pi)^{-\frac{3}{2}}\int
e^{-i\bk\cdot\bx}f(\bx)d^3x.
\end{equation}
Let $U(t)=e^{-iHt}$. Since $H$ is self-adjoint on the domain D($H$),
$U(t)$ is a strongly continuous one-parameter unitary group on
$\ltr$. Let $\phi\in$D($H$). Then $\phi_t\equiv U(t)\phi\in$D($H$)
and satisfies the Schr\"odinger equation
\begin{equation*}
  i\frac{\partial}{\partial   t}\phi_t(\bx)=H\phi_t.
\end{equation*}

In a typical scattering experiment the scattered particles move
almost freely far away from the scattering center. ``Far away" in
position space can also be phrased as ``long before" and ``long
after" the scattering event takes place. So for the ``scattering
states" $\psi$ there are asymptotes
$\psi_{\text{in}},\psi_{\text{out}}$ defined by
\begin{equation}\label{asymptotes}
\begin{split}
  \lim\limits_{t\to -\infty}\|e^{-iH_0t}\psi_{\text{in}}(\bx)-e^{-iHt}\psi(\bx)\|=0\\
\lim\limits_{t\to\infty}\|e^{-iH_0t}\psi_{\text{out}}(\bx)-e^{-iHt}\psi(\bx)\|=0.
\end{split}
\end{equation}
From this it is natural to define the wave operators
$\Omega_{\pm}:\ltr\to\ran(\Omega_{\pm})$ by the strong limits
\begin{equation}\label{waveoperators}
\Omega_{\pm}:=\underset{t\to\pm\infty}{\operatorname{s\:-\:lim}}\;e^{iHt}e^{-iH_0t}.
\end{equation}
These wave operators map the incoming and outgoing asymptotes to
their corresponding scattering states. Ikebe \cite{ikebe:60}
proved that for a potential $V\in (V)_n$ the wave operators exist
and have the range
\begin{equation*}
\text{Ran}(\Omega_{\pm})=\mathcal{H}_{\text{cont}}(H)=\mathcal{H}_{\text{a.c.}}(H).
\end{equation*}
(This property is called asymptotic completeness.) Hence, the
scattering states consist of states with absolutely continuous
spectrum and the singular continuous spectrum of $H$ is empty. In
addition Ikebe \cite{ikebe:60} showed that the Hamiltonian has no
positive eigenvalues. Then we have for every
$\psi\in\mathcal{H}_{\text{a.c.}}(H)$ asymptotes $\pin,\pout\in
\ltr$ with
\begin{equation}\label{defout}
  \Omega_-\psi_{\text{in}}=\psi=\Omega_+\psi_{\text{out}}.
\end{equation}
On D($H_0$) the wave operators satisfy  the so-called intertwining property
\begin{equation*}
 H\Omega_{\pm}=\Omega_{\pm}H_0,
\end{equation*}
while on
$\mathcal{H}_{\text{a.c.}}(H)\cap$D($H$) we have  that
\begin{equation}\label{inter}
 H_0\Omega_{\pm}^{-1}=\Omega_{\pm}^{-1}H.
\end{equation}
The scattering operator $S:\ltr\to\ltr$ is given by
\begin{equation*}
 S:=\Omega_{+}^{-1}\Omega_{-},
\end{equation*}
while using the identity $I$, the $T$-operator is given by
\begin{equation}\label{toperator}
 T:=S-I.
\end{equation}
If the system is asymptotically complete, the ranges of the wave
operators are equal and thus $S$ is unitary. Since the wave
operator maps a scattering state onto its asymptotic state, the
scattering operator maps the incoming asymptote $\pin$ onto the
corresponding out state $\pout.$ The formula for the $T$-matrix,
which holds in the $L^2$-sense, is given by (see e.g., Theorem
XI.42 in \cite{simon3:79})
\begin{equation}\label{formulat}
  \widehat{Tg}(\bk)=-2\pi i\int\limits_{k'=k}T(\bk,\bkp)
\widehat{g}(\bkp)k'd\Omega',
\end{equation}
for $g\in\schwarz$ (Schwartz space) such that $\widehat{g}$ has
support in a spherical shell.\footnote{\zt{In \cite{simon3:79} Equation (\ref{formulat}) was proven outside an ``exceptional set". For our class of potentials the ``exceptional set" is empty.} The additional factor $\frac{1}{2}$ in \cite{simon3:79}
comes from the different definition of $H_0$.} $T(\bk,\bkp)$ is
given by (see e.g., \cite{simon3:79}):
\begin{equation}\label{tmatrix}
  T(\bk,\bkp)=(2\pi)^{-3}\int
e^{-i\bk\cdot\bx}V(\bx)\varphi_-(\bx,\bkp)d^3x,
\end{equation}
where $\varphi_-$ (as well as $\varphi_+$) are eigenfunctions of
$H$ defined by Lemma \ref{propexpansion} below. Since the
eigenfunctions $\varphi_{\pm}$ are bounded and continuous (cf.
Lemma \ref{proplippschw}), we can conclude that $T(\bk,\bkp)$ is
bounded and continuous on $\RRR^3\times\RRR^3,$ if the potential
is in $(V)_3.$ Then the formula (\ref{formulat}) can be proved for
$g\in\schwarz$ without any restriction on the momentum support
by the same method as in \cite{simon3:79}.

We will need the time evolution of a state
$\psi\in\mathcal{H}_{\text{a.c.}}(H)$ with the Hamiltonian $H$. Its
diagonalization on $\mathcal{H}_{\text{a.c.}}(H)$ is given by the
eigenfunctions $\varphi_{\pm}$\label{lipp}:
\begin{equation}\label{schrlipp}
(-\frac{1}{2}\Delta+V(\bx))\varphi_{\pm}(\bx,\bk)=\frac{k^2}{2}\varphi_{\pm}(\bx,\bk).\end{equation}
Inverting $(-\frac{1}{2}\Delta-\frac{k^2}{2})$ one obtains the
Lippmann-Schwinger equation. We recall the main parts of a result
on this due to Ikebe in \cite{ikebe:60} which is collected in the
present form in \cite{teufel2:99}.

\bprop\label{propexpansion} Let $V\in (V)_2$. Then for any
$\bk\in\RRR^3\backslash\{0\}$ there are unique solutions
$\varphi_{\pm}(\cdot,\bk):\RRR^3\to\mathbb{C}$ of the
Lippmann-Schwinger equations
\begin{equation}\label{lippmann}
\varphi_{\pm}(\bx,\bk)=e^{i\bk\cdot\bx}-\frac{1}{2\pi}\int\frac{e^{\mp
  ik|\bx-\bxp|}}{|\bx-\bxp|}V(\bxp)\varphi_{\pm}(\bxp,\bk)d^3x',
\end{equation}
which satisfy the boundary conditions
$\lim_{|\bx|\to\infty}(\varphi_{\pm}(\bx,\bk)-e^{i\bk\cdot\bx})=0$,
which are also classical solutions of the stationary Schr\"odinger
equation (\ref{schrlipp}), and are such that:
\begin{itemize}
\item[(i)] For any $f\in \ltr$ the generalized Fourier
transforms\footnote{$\limmean\int$ is a shorthand notation for
$\underset{R\to\infty}{\operatorname{s-lim}}\;\int_{B_R}$, where $\text{s\:-}\lim$ denotes the limit in the $L^2$-norm and $B_R$ a ball with radius $R$ around the origin.}

\begin{equation*}
  (\mathcal{F}_{\pm}f)(\bk)=\frac{1}{(2\pi)^{\frac{3}{2}}}\limmean  
\int\varphi_{\pm}^{\ast}(\bx,\bk)f(\bx)d^3x
\end{equation*}
exist in $\ltr$.
\item[(ii)] Ran($\mathcal{F}_{\pm})=\ltr.$ Moreover $\mathcal{F}_{\pm}: \mathcal{H}_{\text{a.c.}}(H) \to \ltr $
are unitary and the inverses of these unitaries are given by
\begin{equation*}
(\mathcal{F}_{\pm}^{-1}f)(\bx)=\frac{1}{(2\pi)^{\frac{3}{2}}}\limmean
  \int\varphi_{\pm}(\bx,\bk)f(\bk)d^3k.
\end{equation*}

\item[(iii)] For any $f\in \ltr$ the relations
$\Omega_{\pm}f=\mathcal{F}_{\pm}^{-1}\mathcal{F}f$ hold, where
$\mathcal{F}$ is the ordinary Fourier transform given by
(\ref{fourier}).
\item[(iv)] For any
$f\in\operatorname{D}(H)\cap\mathcal{H}_{\text{a.c.}}(H)$ we
have:
\begin{equation*}
Hf(\bx)=\left(\mathcal{F}_{\pm}^{-1}\frac{k^2}{2}\mathcal{F}_{\pm}f\right)(\bx)
\end{equation*}
and therefore for any $f\in\mathcal{H}_{\text{a.c.}}(H)$
\begin{equation*}
e^{-iHt}f(\bx)=\left(\mathcal{F}_{\pm}^{-1}e^{-i\frac{k^2}{2}t}\mathcal{F}_{\pm}f\right)(\bx).
\end{equation*}

\end{itemize}
\eprop

In order to apply stationary phase methods we will need estimates on
the derivatives of the generalized eigenfunctions:

\bprop\label{proplippschw} Let $V\in(V)_n$ for some $n\geq 3.$ Then:
\begin{itemize}

\item[(i)]$\varphi_{\pm}(\bx,\cdot)\in
C^{n-2}(\RRR^3\setminus\{0\})$ for all $\bx\in\RRR^3$ and the
partial derivatives\footnote{We use the usual multi-index notation:
$\alpha=(\alpha_1,\alpha_2,\alpha_3),\;\alpha_i\in\NNN_0,\;\partial_{\bk}^{\alpha}f(\bk):=\partial_{k_1}^{\alpha_1}\partial_{k_2}^{\alpha_2}\partial_{k_3}^{\alpha_3}f(\bk)\text{
and }|\alpha|:=\alpha_1+\alpha_2+\alpha_3$.}

$\partial_{\bk}^\alpha\varphi_{\pm}(\bx,\bk),$ $|\alpha|\leq n-2,$
are continuous with respect to $\bx$ and $\bk.$
\end{itemize}
If, in addition, zero is neither an eigenvalue nor a resonance of
$H$, then
\begin{itemize}

\item[(ii)]$\sup\limits_{\bx\in\mathbb{R}^3,\bk\in\mathbb{R}^3}|\varphi_{\pm}(\bx,\bk)|<\infty,$
\end{itemize}
for any $\alpha$ with $|\alpha|\leq n-2$ there is a
$c_{\alpha}<\infty$ such that
\begin{itemize}

\item[(iii)]$\sup\limits_{\bk\in\ \mathbb{R}^3\setminus\{0\}}
|\kfac^{|\alpha|-1}\partial_{\bk}^\alpha\varphi_{\pm}(\bx,\bk)| <
c_\alpha\lxr^{|\alpha|},\;\text{ with }\kfac:=\frac{k}{\lkr},$
\end{itemize}
and for any $l\in \{1,...,n-2\}$ there is a
$c_{l}<\infty$ such that
\begin{itemize}

\item[(iv)]$\sup\limits_{\bk\in\ \mathbb{R}^3\setminus\{0\}}
\left|\frac{\partial^l}{\partial k^l}\varphi_{\pm}(\bx,\bk)\right|
< c_l\lxr^{l},$ where $\frac{\partial}{\partial k}$ is the radial
partial derivative in $\bk$-space.
\end{itemize}

\eprop

\brem This proposition, except the assertion (iii), was proved in
\cite{teufel2:99}, Theorem 3.1. Assertion (iii) repairs a false
statement in Theorem 3.1 which did not include the necessary
$\kfac^{|\alpha|-1}$ factor, which we have in (iii). For
$|\alpha|=1,$ which was the important case in that paper, there is
however no difference. We have commented on the  proof of this
corrected version  in \cite{duerr1:04}. \erem

\brem Zero is a resonance of $H$ if there exists a solution $f$ of
$Hf=0$ such that $\lxr^{-\gamma}f\in\ltr$ for any
$\gamma>\frac{1}{2}$ but not for $\gamma=0.$\footnote{There are
various definitions, see e.g. \cite{yajima:95}, p. 552,
\cite{albeverio:88}, p.20 and \cite{jensen:79}, p. 584.} The
appearance of a zero eigenvalue or resonance can be regarded as an
exceptional event: For a Hamiltonian $H=H_0+cV,\;c\in\RRR,$ this
can only happen for $c$ in a discrete subset of $\RRR,$ see
\cite{albeverio:88}, p. 20 and \cite{jensen:79}, p. 589.\erem

As a simple consequence of Proposition \ref{proplippschw} we
obtain

\bcor\label{cort} Let $V\in(V)_3$ and let zero be neither an eigenvalue nor a resonance of $H$. Then the $T$-matrix defined by (\ref{tmatrix}) is a bounded and continuous function on $\RRR^3\times\RRR^3.$ Moreover, if $V\in(V)_n$, for some $n\geq 3$ we have for all multi-indices $\alpha$ with $|\alpha|\leq n-3$ a constant $c_\alpha>0$ such that
\begin{align}
\sup\limits_{\bkp\in\RRR^3,\bk\in\RRR^3\setminus\{0\}}\kfac^{|\alpha|-1}|\partial_{\bk}^{\alpha}T(\bkp,\bk)|\leq c_\alpha.
\end{align}
\ecor

With the regularity of the generalized eigenfunctions one can
prove the flux-across-surfaces theorem (FAST). The quantum probability
current density (=quantum flux density) is given by
\begin{equation}\label{psiflux}
  \bj^{\psi_t}(\bx):=-\frac{i}{2}(\psi_t^\ast(\bx)\nabla
\psi_t(\bx)-\psi_t(\bx)\nabla   \psi_t^\ast(\bx)).
\end{equation}
For $\psi_t(\bx)$ a solution of the Schr\"odinger equation we have
the identity
\begin{equation*}
  \frac{\partial |\psi_t(\bx)|^2}{\partial t}+\diver
\bj^{\psi_t}(\bx)=0,
\end{equation*}
which has the form of a continuity equation. The
flux-across-surfaces theorem can be naturally proven for the
following class of wave functions (in the following definition we
have the Fourier transform of $\pout$, $\pouth(\bk)=\int\varphi_+(\bx,\bk)\psi(\bx)d^3x$ (cf. Proposition
\ref{propexpansion}), in mind):

\bde\label{defg} A function $f:\RRR^3\setminus\{0\}\to\CCC$ is in $\ghut$
if there is a constant $C\in\RRR_+$ with:
\begin{equation*}|f(\bk)|\leq C\lkr^{-15},\end{equation*} \begin{equation*}\left|\partial^{\alpha}_{\bk}f(\bk)\right|\leq C\lkr^{-6},\;|\alpha|=1,\end{equation*} \begin{equation*}\left|\kfac\:\partial^{\alpha}_{\bk}f(\bk)\right|\leq C\lkr^{-5},\;|\alpha|=2,\;\kfac=\frac{k}{\lkr},\end{equation*}  \begin{equation*}\left|\frac{\partial^2}{\partial k^2}f(\bk)\right|\leq C\lkr^{-3}.\end{equation*} \ede

With this definition we have

\bprop\label{propflux}\hspace{-0,25cm}(Flux-across-surfaces
theorem)\hspace{0,1cm} Suppose $V\in(V)_4$ and that zero is
neither a resonance nor an eigenvalue of $H$. Suppose
$\pouth(\bk)\in\ghut$ and let
$\psi=\Omega_+\pout.$ Then $\psi_t(\bx)=e^{-iHt}\psi(\bx)$ is
continuously differentiable except at the singularities of $V$,
for any measurable set $\Sigma\subseteq S^2$ and any
$T\in\mathbb{R}$ $\bj^{\psi_t}(\bx)   \cdot d\bsig dt$ is
absolutely integrable on $R\Sigma\times[T,\infty)$ for R
sufficiently large and
\begin{align}\label{flux}
\lim\limits_{R\to\infty}\int\limits_T^{\infty}\int\limits_{R\Sigma}\bj^{\psi_t}(\bx)
  \cdot d\bsig dt&=  
\lim\limits_{R\to\infty}\int\limits_T^{\infty}\int\limits_{R\Sigma}\left
|\bj^{\psi_t}(\bx)   \cdot d\bsig\right|
dt=\int\limits_{C_{\Sigma}}|\pouth(\bk)|^2d^3k,
\end{align}
where $R\Sigma:=\{\bx\in\RRR^3:\bx=R\bom,\;\bom\in\Sigma\}$,
$C_{\Sigma}:=\{\bk\in\RRR^3:\frac{\bk}{k}\in\Sigma\}$ is the cone
given by $\Sigma$ and $d\bsig$ is the outward-directed surface
element on $RS^2.$ \eprop The proof can be found in
\cite{duerr1:04}.

The FAST plays a crucial role in the proof of our main results,
Theorem \ref{thedasisses} and Theorem \ref{thedasisseslight}. Its
importance for scattering theory was first pointed out in
\cite{combes:75}.

\section{The quantum flux, crossing statistics and Bohmian mechanics}\label{chflux}\sectionmark{Quantum flux and crossing statistics}

In Bohmian mechanics, see \cite{bohm:52}, the particle has a position $\bbq_t$ that evolves via the
equations
\begin{equation}\label{bohm}
\begin{split}
\frac{d}{dt}\bbq_t=\bv^{\psi_t}(\bbq_t)=\operatorname{Im}\frac{\nabla
  \psi_t}{\psi_t}(\bbq_t),\\
  i\frac{\partial}{\partial t}\psi_t(\bx)=H\psi_t(\bx).
\end{split}
\end{equation}
According to the quantum equilibrium hypothesis (\cite{duerr:92},
Born's law), the positions of particles in an ensemble of
particles each having wave function $\psi$ are always
$|\psi|^2$-distributed. Note that if $\bbq_0$ is
$|\psi_0|^2$-distributed then $\bbq_t$ is
$|\psi_t|^2$-distributed.

Under two assumptions we have the $|\psi_0|^2$ almost-sure existence
and uniqueness of the Bohmian dynamics:

\bcond\label{condcyc}The initial wave function $\psi_0$ is
normalized, $\|\psi_0\|=1$, and $\psi_0\in
C^{\infty}(H)=\bigcap\limits_{n=1}^{\infty}D(H^n).$\econd

\bcond\label{condpot}The potential $V$ is in $V_2$ and
$C^{\infty}$ except, perhaps, at a finite number of
singularities.\econd (See Berndl et al. \cite{berndl:95}, Theorem
3.1 and Corollary 3.2 for the proof, as well as Theorem 3 and
Corollary 4 in \cite{teufel:04}. The conditions in
\cite{berndl:95,teufel:04} are much more general. In our context,
however, we have to restrict to the case where $V\in(V)_2$.)
Hence, depending on the initial position $\bq_0\in\Omega_0$, where
$\Omega_0$ is the set of ``good'' points, the particle has the
trajectory $\bbq_t^{\psi}(\bq_0).$ On the set of ``good'' points,
$\psi_0(\bx)$ is different from zero and is differentiable. The
complement $\RRR^3\setminus\Omega_0$ of $\Omega_0$ has measure 0
(with respect to $|\psi_0|^2$).

Given a trajectory $\bbq_t^{\psi}(\bq_0)$, $\bq_0\in\Omega_0$, we
can define the number of crossings in a natural way. For the surface
$R\Sigma\subset RS^2$ with unit and normal vector $\bn(\bx)=\frac{\bx}{x},\;\bx\in R\Sigma$ we define $N_+^{\psi}(R\Sigma)$ on $\Omega_0$ by:
\begin{equation}\label{nplus}
 N_+^{\psi}(R\Sigma)(\bq_0):=\left|\left\{t\geq0|\bbq_t^{\psi}(\bq_0)\in R\Sigma\text{ and
}\dot{\bbq}_t^{\psi}(\bq_0)\cdot\bn\left(\bbq_t^{\psi}(\bq_0)\right)>0\right\}\right|,
\end{equation}
the number of crossings of the trajectory $\bbq^{\psi}_t(\bq_0)$
through $R\Sigma$ in the direction of the orientation in the time
interval $[0,\infty)$ (``problematical crossings'' where the
velocity is ``orthogonal'' to the orientation of $R\Sigma$ have
measure zero and need not concern us, see \cite{berndl:94}, p.
28-34). If $N_+^{\psi}(R\Sigma)(\bq_0)\geq 1$, we can define
$t_{\text{exit}}^{R\Sigma}$ as the time  when the particle crosses
the surface $R\Sigma$ in the positive direction for the first
time:
\begin{equation}\label{texit}
 t_{\text{exit}}^{R\Sigma}(\bq_0):=\min\left\{t\geq
0|\bbq_t^{\psi}(\bq_0)\in R\Sigma\text{ and
}\dot{\bbq}_t^{\psi}(\bq_0)\cdot\bn\left(\bbq_t^{\psi}(\bq_0)\right)>0\right\}.
\end{equation}
In the case that the particle does not cross the surface in the
positive direction, we set
\begin{equation}\label{texits}
 t_{\text{exit}}^{R\Sigma}(\bq_0):=\infty,\text{ if
}N_+^{\psi}(R\Sigma)(\bq_0)=0.
\end{equation}
Analogously to (\ref{nplus}) we have $N_-^{\psi}(R\Sigma),$ the
number of crossings in the opposite direction. For convenience we
define $N_+^{\psi}(R\Sigma)$ and $N_-^{\psi}(R\Sigma)$ on the
whole of $\RRR^3$ by setting
$N_+^{\psi}(R\Sigma)=N_-^{\psi}(R\Sigma)=0$ for all
$\bq_0\in\RRR^3\setminus\Omega_0.$ Then we can define the number
of signed crossings on $\RRR^3$ by
\begin{equation}
  \label{defnsig} N_{\text{sig}}^{\psi}(R\Sigma):=N_+^{\psi}(R\Sigma)-N_-^{\psi}(R\Sigma).
\end{equation}
The total number of crossings defined on $\RRR^3$ is then
\begin{equation}
  \label{defntot} N_{\text{tot}}^{\psi}(R\Sigma):=N_+^{\psi}(R\Sigma)+N_-^{\psi}(R\Sigma).
\end{equation}
These quantities are random variables on the space $\RRR^3$ of
initial conditions, see \cite{berndl:94}, Lemma 4.2. The
expectation values of $N_{\text{sig}}^{\psi}(R\Sigma)$ and
$N_{\text{tot}}^{\psi}(R\Sigma)$ are given by flux integrals and
are finite, see Proposition \ref{propexpn} below. This means that
$N_{\text{sig}}^{\psi}(R\Sigma)$ and
$N_{\text{tot}}^{\psi}(R\Sigma)$ are almost surely finite. Before
we give a precise statement we argue  heuristically for the
connection between the quantum flux and the expectation values.
 For a particle to cross an infinitesimal surface $d\bsig:=\bn
d\sigma$ in a time interval $[t,t+dt)$, it must be at time $t$ in
the appropriate cylinder of size $|\bv^{\psi_t}(\bx)\cdot d\bsig
dt|$. The probability is therefore
\begin{equation*}
 |\psi_t(\bx)|^2|\bv^{\psi_t}(\bx)\cdot d\bsig
 dt|=|\bj^{\psi_t}(\bx)\cdot d\bsig|dt.
\end{equation*}
Because the intervals are infinitesimal, we have for
$N_{\text{sig}}^{\psi}(dt,d\bsig)\in\{-1,0,1\},$\footnote{$N_{\text{sig}}^{\psi}(dt,d\bsig)$ is the
number of signed crossings in the time interval $[t,t+dt)$ through
the surface $d\bsig$.} where the sign will be the same as that of
$\bj\cdot d\bsig.$  Therefore
$\mathbb{E}(N_{\text{sig}}^{\psi}(dt,d\bsig))=\bj^{\psi_t}(\bx)\cdot
d\bsig dt$ and integration over $R\Sigma$ and $[0,\infty)$ yields
(\ref{expnsig}). The precise statement is:

\bprop\label{propexpn} Let A\ref{condcyc} and A\ref{condpot} be
satisfied. In addition suppose that the conditions of Proposition \ref{propflux} are satisfied. Then for sufficiently large $R$ the expectation values of $N_{\text{sig}}^{\psi}(R\Sigma)$ and $N_{\text{tot}}^{\psi}(R\Sigma)$ are finite and
\begin{equation}\label{expnsig}
\mathbb{E}(N_{\text{sig}}^{\psi}(R\Sigma))=\int\limits_{0}^{\infty}\int\limits_{R\Sigma}\bj^{\psi_t}(\bx)\cdot
d\bsig dt,
\end{equation}

\begin{equation}\label{expn}
\mathbb{E}(N_{\text{tot}}^{\psi}(R\Sigma))=\int\limits_{0}^{\infty}\int\limits_{R\Sigma}|\bj^{\psi_t}(\bx)\cdot
d\bsig| dt.
\end{equation}

\eprop
The proof of Proposition \ref{propexpn} can be found in
\cite{berndl:94}, pp. 34-37, and under slightly different
conditions in \cite{tumulka:01}. The results in the references
hold under more general conditions on the surfaces.

Consider now a scattering situation where we want to calculate the number of first crossings. The detector corresponds to
the surface
$R\Sigma:=\{\bx\in\RRR^3:\bx=R\bom,\;\bom\in\Sigma\subset
S^2\}\subset RS^2$. Then we define $\nd^{\psi}(\dti,R,\Sigma)$ to be
equal to one if the particle with the wave function $\psi_0=\psi$ is
``detected'' in $\dti$ and zero otherwise. More precisely,\newpage
\begin{equation*}
\nd^{\psi}(R,\Sigma):\RRR^3\to\{0,1\},
\end{equation*}
\begin{equation}\label{ndetpsi}
\nd^{\psi}(R,\Sigma)(\bq_0):=   \begin{cases}
 1,\text{ if }q_0\leq R, t_{\text{exit}}^{ RS^2}<\infty\text{ and }\bbq_{t_{\text{exit}}^{ RS^2}}^{\psi}(\bq_0)\in R\Sigma,\\
    0\text{ otherwise}.   \end{cases}  
\end{equation}
The definition is motivated by the idea that particles are
detected when they cross the boundary $ RS^2$ for the first time.
Using the fact that $ RS^2$ is closed we can estimate
\begin{equation*}
\left|\nd^{\psi}(R,\Sigma)-N_{\text{sig}}^{\psi}(R\Sigma)\right|\leq
N_-^{\psi}(RS^2)
\end{equation*}
so that by the triangle inequality
\begin{equation}\label{uniform}
 \left|\mathbb{E}(\nd^{\psi}(R,\Sigma))-
\mathbb{E}(N_{\text{sig}}^{\psi}(R\Sigma))\right|\leq
\mathbb{E}   (N_-^{\psi}( RS^2)).
\end{equation}
With (\ref{defnsig}), (\ref{defntot}) and Proposition
\ref{propexpn} we obtain for the right-hand side of
(\ref{uniform})
\begin{align}\label{unifor}
 \mathbb{E}   (N_-^{\psi}(RS^2))&=\frac{1}{2}\mathbb{E}\left(N_{\text{tot}}^{\psi}(RS^2)-N_{\text{sig}}^{\psi}(RS^2)\right)=\frac{1}{2}\int\limits_0^{\infty}\int\limits_{RS^2} \left(|\bj^{\psi_t}(\bx)\cdot d\bsig|-\bj^{\psi_t}(\bx)\cdot
d\bsig\right) dt.
\end{align}
If $\bj^{\psi_t}(\bx)\cdot d\bsig\geq 0$ for all
$d\bsig\in RS^2$ and $t>0$ then we have by (\ref{uniform})
and (\ref{unifor}) that
$\mathbb{E}(N_{\text{sig}}^{\psi}(R\Sigma))=\mathbb{E}(\nd^{\psi}(R\Sigma)).$
In general $\bj^{\psi_t}(\bx)\cdot d\bsig$ does not have to be
positive, but the flux-across-surfaces theorem (Proposition
\ref{propflux}) ensures that the flux is asymptotically outwards.
Thus we can estimate the difference between
$\mathbb{E}(N_{\text{sig}}^{\psi}(R\Sigma))$ and
$\mathbb{E}(\nd^{\psi}(R\Sigma))$ for all $\psi$ which satisfy the
flux-across-surfaces theorem using (\ref{uniform}) and
(\ref{unifor})
\begin{align}\label{uni1}
\left|\mathbb{E}(N_{\text{sig}}^{\psi}(R\Sigma))-\mathbb{E}(\nd^{\psi}(R,\Sigma))\right|\leq\frac{1}{2}\int\limits_0^{\infty}\int\limits_{RS^2} \left(|\bj^{\psi_t}(\bx)\cdot d\bsig|-\bj^{\psi_t}(\bx)\cdot
d\bsig\right) dt\underset{R\to\infty}{\to}0.
\end{align}
In particular under the hypotheses of Proposition \ref{propflux}
and the general assumptions A\ref{condcyc} and A\ref{condpot} we
obtain asymptotic equality between the expectation values
$\mathbb{E}(\nd^{\psi}(R,\Sigma))$ and
$\mathbb{E}(N_{\text{sig}}^{\psi}(R\Sigma))$.

\section{A model for the beam}\label{chmodel}
In a scattering situation a beam of particles is scattered off a target. We now wish to focus on the beam. We take the beam to be produced by a particle source located in the plane $\GammaL$\label{impact}
perpendicular to the $x_3$-axis:
\begin{equation*}
  \GammaL:=\{-L\be_3+\ba|\;\ba\bot\be_3\},\;L>0.
\end{equation*}

The particles are created with wave functions
$\psi\in\mathcal{H}_{\text{a.c.}}$ translated to the plane
$\GammaL$. Calling $\psi_{\by}$ the translation of $\psi$ by
$\by$, the ``centers'' of the translated wave functions, with
which we are concerned, are located at
\begin{equation*}
  \by=y_1\be_1+y_2\be_2-L\be_3\in\GammaL
\end{equation*}
and are uniformly \zt{distributed} in a bounded region
$A\subset\GammaL$ with area $|A|$. We call $A$ the beam profile.
The momentum distribution of the wave function is concentrated
around the momentum $\bko\|\be_3$.

\brem This model of a beam, in which the particles have random
impact parameters and are scattered off a single target
``particle,'' is equivalent to the more realistic description of the
scattering situation, in which all the target particles are randomly
distributed (e.g., in a foil) and the incoming particles have the
very same impact parameter, provided coherent and
multiple-scattering effects are neglected (see e.g. \cite{newton:82},
p. 214).\erem

The translated wave function $\psi_{\by}$ of a wave function
$\psi\in\mathcal{H}_{\text{a.c.}}$ will not in general be in
$\mathcal{H}_{\text{a.c.}}$, but can have a part in
$\mathcal{H}_{\text{p.p.}}.$ This is problematical for the
application of our general results (see Section \ref{chproblem}). To
avoid this difficulty, we assume:

\bcond\label{condbound} The Hamiltonian $H=-\frac{1}{2}\Delta+V$
has no bound states, i.e.
$\mathcal{H}_{\text{p.p.}}=\{0\}.$\econd
Then
$\psi_{\by}\in\mathcal{H}_{\text{a.c.}},\forall\by\in\RRR^3.$

\zt{We specify now more precisely the model for the beam, which has been already mentioned in \cite{duerr:00}.} The particles are created with wave functions $\psi$ at random times $t\in\RRR_+$ and where the wave
function of a particle is shifted randomly by the uniformly
distributed ``impact parameter'' $\by\in A$, the ``center" of the
wave function at the moment of emission. In Bohmian mechanics the
initial position $\bq\in\RRR^3$ of the particle determines its
trajectory. The initial position is
$|\psi_{\by}|^2$-distributed. We shall not need many stochastic
details about the beam. The reader may think of
a Poisson point process with points in 
\begin{equation*}\Lambda=\mathbb{R}^+\times
A\times\RRR^3,
\end{equation*} 
with a point $\lambda=(t,\by,\bq)\in\Lambda$ representing a particle with
wave function
\begin{equation}\label{psi}
  \psi_{\by}(\bx)\equiv\psi(\bx-\by),\;\by\in A
\end{equation}
emitted at the time $t\in\RRR^+$ and with initial position
$\bq\in\RRR^3.$ We shall consider a general point process
$(\Lambdastar,\mathfrak{F},\mathbb{P})$ built on
$(\Lambda,\mathfrak{B}(\Lambda),\mu)$, where
$\lambdastar\in\Lambdastar$ represents a configuration of
countably many points in $\Lambda$, i.e.
\begin{equation*}
\lambdastar=\{\lambda\},\;\lambda\in\Lambda,\;\lambdastar\text{ countable.}
\end{equation*}
For the number of points
\begin{equation*}
\chara^{\star}(\lambdastar)\equiv\sum\limits_{\lambda\in\lambdastar}\chara(\lambda)
\end{equation*}
in a set $B\in\mathfrak{B}(\Lambda)$, where $\chara$ is the
indicator function of the set $B$, we have that
\begin{equation}\label{expnb}
\mathbb{E}\left(\chara^{\star}\right)=\mu(B),
\end{equation}
where the intensity measure $\mu$ on $\mathfrak{B}(\Lambda)$ is
given by
\begin{equation}\label{mu}
 d\mu=|\psi(\bx-\by)|^2\chi_{A}(\by)dtd^2yd^3x.
\end{equation}

\brem\label{rempoisson1} For a Poisson process we would have, in
addition to (\ref{expnb}), that
\begin{equation}\label{poissonp}
\mathbb{P}\left(\chara^{\star}=k\right)=\exp(-\mu(B))\frac{\mu(B)^k}{k!}
\end{equation}
as well as the \z{independence of $\chi_A^{\star}$ and $\chara^{\star}$,} for
$A\cap B=\emptyset,\;A,B\in\mathfrak{B}(\Lambda).$ \erem

We shall assume that the point process is ergodic in the following
sense: For any $B\in\mathfrak{B}(\Lambda)$ let
\begin{equation}\label{bdte}
B(\dte):=\{(t,\by,\bq)\in B|t\in[0,\dte)\}.
\end{equation}
Then for any $\epsilon>0$
\begin{equation}\label{lawlarge} 
\lim\limits_{\dteb\to\infty}\mathbb{P}\left(\left|\frac{\charac_{B(\dte)}^{\star}}{\dteb}-\mathbb{E}\left(\frac{\charac_{B(\dte)}^{\star}}{\dteb}\right)\right|\geq\epsilon\right)=0,
\end{equation}
with $\mathbb{E}\left(\charac_{B(\dte)}^{\star}\right)$ given
by (\ref{expnb}).

\brem Because of the independence property (cf. Remark
\ref{rempoisson1}), (\ref{lawlarge}) holds for the case of
a Poisson process.\erem

\brem\label{remdens} \zt{The point process has unit density in the following sense: Let} $\Ad\subset A$, $\dteb>0$ and $B:=[0,\dteb)\times \Ad\times\RRR^3$ be given. Then with (\ref{lawlarge}) for any $\epsilon>0$
\begin{equation}\label{dens1} 
\lim\limits_{\dteb\to\infty}\mathbb{P}\left(\left|\frac{\charac_{B}^{\star}}{|\Ad|\dteb}-\mathbb{E}\left(\frac{\charac_{B}^{\star}}{|\Ad|\dteb}\right)\right|\geq\epsilon\right)=0,
\end{equation}
and 
\begin{equation}\label{dens2} 
\mathbb{E}\left(\frac{\charac_{B(\dte)}^{\star}}{|\Ad|\dteb}\right)=\frac{1}{|\Ad|\dteb}\mu(B)=1.
\end{equation}
\erem

\section{The definition of the scattering cross section}\label{chdefcross}

We shall now start to define $\ndstar(\dte,R,A,L,\psi,\Sigma)$, the
number of detected particles. To simplify the notation we do not
always \z{indicate} the dependence of $\ndstar$ on $A,L$ and $\psi$.
Sometimes we will also suppress the dependence on $R$ and $\Sigma$.
We define first $\nd(\dte,R,\Sigma)$ for a single particle corresponding
to $\lambda=(t,\by,\bq)$ by
\begin{equation*}
\nd(\dte,R,\psi,\Sigma):\Lambda\to\{0,1\},
\end{equation*}
\begin{equation}\label{ndetl}
\nd(\dte,R,\psi,\Sigma)(\lambda):=\chi_{[0,\dte)}(t)\nd^{\psi_{\by}}(R,\Sigma)(\bq),
  \end{equation} where $\nd^{\psi_{\by}}(R,\Sigma)(\bq)$ is
defined by (\ref{ndetpsi}). The characteristic function ensures
that no particle is counted which is emitted after the time $\dte.$ Note that $\psi_{\by}$ must satisfy condition
A\ref{condcyc} (p. \pageref{condcyc}) to ensure that
$\nd^{\psi_{\by}}(R,\Sigma)(\bq)$ is well defined. Then
\begin{equation*}
\ndstar(\dte,R,A,L,\psi,\Sigma):\Lambdastar\to\mathbb{N}_0,
\end{equation*}
\begin{equation}\label{ndet}
\ndstar(\dte,R,A,L,\psi,\Sigma)(\lambdastar)=\sum\limits_{\lambda\in\lambdastar}\nd(\dte,R,\psi,\Sigma)(\lambda).
\end{equation}

\zt{The empirical scattering cross section
$\sigma_{\text{emp}}(\Sigma)$ for the solid angle $\Sigma$ is the} \z{random variable\footnote{\zt{We shall ignore the dimension factor $[\text{unit area}\cdot\text{unit
time}]$ which comes from the normalization of (\ref{def}) by the unit density $\frac{1}{[\text{unit area}\cdot\text{unit
time}]}$ of the underlying point process, cf. Remark \ref{remdens}. One can also normalize by the beam density, i.e. with the number of detected particles (by a detector in the beam with a surface perpendicular to the beam axis) per unit time and unit area, in front of the target. In the scattering regime, i.e. if the parameters are suitably scaled (cf. Section \ref{chregime}), the beam will have unit density in front of the target. We shall not elaborate on this further in this paper, see however \cite{duerr4:04}.}} 
\begin{equation}\label{def}
  \sigma_{\rm emp}(\Sigma):=\frac{\ndstar(\dte,R,A,L,\psi,\Sigma)}{\dteb},
\end{equation}
which} by the law of large numbers (for
the Poisson case and by the ergodicity assumption (\ref{lawlarge})
for the general case) should approximate for large $\dteb$ in
$\mathbb{P}$-probability its corresponding
$\mathbb{P}$-expectation value. The expected value of (\ref{def})
is then the theoretically predicted cross section. This
theoretically predicted cross section involves a very complicated
formula which is not very explicit, cf. (\ref{gamma}) and Remark
\ref{rempoisson}. It depends of course on the detection \z{directions}
$\Sigma$, the potential $V$ and the \z{approximate} momentum $\bko$ of the
particles in the beam, but depends also on the other details of
the experimental setup such as $R$, $A$, $L$ and the detailed
specification of $\psi$. By taking the scaling limit described in
the next \z{section,} we shall arrive at (\ref{rums}), which does not
depend on these additional details.

\section{The scaling of the parameters}\label{chregime}

According to the usual asymptotic picture of scattering theory where the
particles are prepared long before and are detected long after the
scattering event has \z{occurred,} the preparation and detection should be
far away from the scattering center. That means the limits $R\to\infty$ and
$L\to\infty$ have to be \z{taken}.  \z{However,} increasing $L$ has the
(\z{undesirable}) effect of an \z{increased} spreading of the beam, which
reduces the beam intensity in the scattering region. To \z{maintain the}
beam intensity in the scattering region \z{we must widen} the beam profile
$A$ \z{as $L\to\infty$.} The \z{idealization of an incoming plane wave
corresponds to} particles with a narrow distribution in momentum space,
\z{i.e., to a limit in which the Fourier transform of the initial wave
function becomes more and more concentrated around a fixed initial wave
vector} $\bko$. \zt{For a detailed discussion of the scattering regime see \cite{duerr4:04}.}

\z{The limits for the parameters $L,A,$ and $\psi$ will be combined by
simultaneously scaling them using a small parameter $\epsilon$\,: We
introduce $\Leps$, $\Aeps$ and $\psieps$, whose precise dependence on
$\epsilon$ will be given below, and consider the cross section
corresponding to (\ref{def}), depending on $\epsilon,R,\dteb$,
\begin{align}\label{defeps}
\sigma_{\text{emp}}^{\epsilon}(\Sigma)=\frac{\ndstar(\dte,R,\Aeps,\Leps,\psieps,\Sigma)}{\dteb},
\end{align}
to which the limit  $\epsilon\to 0$ is to be applied.} 

\z{However, the limit $R\to\infty$ is taken before we take $\epsilon\to
0$; this is because we must have that the diameter of the beam profile $A$
is much smaller than $R$, since otherwise unscattered particles will often
contribute to what should be the cross section for scattered particles. For
convenience, we first take the limit $\dteb\to\infty$, required for the
stabilization of the empirical cross section produced by the law of large
numbers.  We are thus led to consider a limit for the cross section of the
form}
\begin{align}\label{defepslim}
\siglim(\Sigma)=\lim\limits_{\epsilon\to
0}\lim\limits_{R\to\infty}\lim\limits_{\dteb\to\infty}\sigma_{\text{emp}}^{\epsilon}(\Sigma).
\end{align}

The precise definition of $\Leps$, $\Aeps$ and $\psieps$, \z{used in our
  main results, is the following:}
\begin{equation}\label{psieps}
 \psieps(\bx)=\epsilon^{\frac{3}{2}}e^{i\bko\cdot\bx}\psi(\epsilon\bx),
\end{equation}
with the Fourier transform
\begin{equation}\label{psiepsfou}
  \widehat{\psieps}(\bk)=\epsilon^{-\frac{3}{2}}\widehat{\psi}\left(\frac{\bk-\bko}{\epsilon}\right).
\end{equation}
The particle source is located \z{on $Y_{\Leps}$, with}
\begin{equation}\label{leps}
 \Leps=\frac{L}{\epsilon^l}, \quad l>2.
\end{equation}
\z{For  the beam profile $\Aeps\subset Y_{\Leps}$ we take the circular  region
\begin{equation}\label{aeps}
 \Aeps=\{\bx\in\RRR^3|\sqrt{x_1^2+x_2^2}<\frac{\Deps}{2}\text{ and
}x_3=\Leps\}
\end{equation}
with the beam  diameter $\Deps$ given by
\begin{equation}\label{deps}
 \Deps=\frac{D}{\epsilon^d}, \quad d>2l-3.
\end{equation}
}
\z{(One might be inclined to consider a scattering experiment in which the
diameter of the beam is much smaller than the distance of the particle
source from the scattering center. \z{Indeed, if $2<l<3$, $d<l$} is
consistent with (\ref{deps}). Hence, such a scenario is covered by our
results.)}

\section{The Scattering Cross Section Theorem}\label{chlimit}

We can now formulate our main results. \z{Our basic assumptions are that
$V\in (V)_5$} (Definition \ref{defpot}), A\ref{condpot} (p.
\pageref{condpot}), A\ref{condbound} (no bound states, p.
\pageref{condbound}) and \z{that for all $\epsilon$ small enough
$\psieps_{\by}$ is ``good'' for all $\by\in\Aeps$ in the sense that it
satisfies A\ref{condcyc} (p. \pageref{condcyc}) as well as the condition
for the FAST (p. \pageref{propflux}).} Moreover, \z{we need to assume that
the potential has no zero energy resonances. However, instead of invoking
the implicit condition on $\psi$ that the $\psieps_{\by}$ are ``good," we
impose stronger but more explicit conditions on $\psi$, namely that
$\psi\in C_0^\infty(\RRR^3)$ (Theorem \ref{thedasisseslight}) or
$\psi\in\mathcal{S}$ (Theorem \ref{thedasisses}), with corresponding
additional conditions on the potential (Definitions \ref{defpotstronglight}
and \ref{defpotstrong}, respectively).}

\bde\label{defpotstrong} $V$ is in $\mathcal{V}$ if
\begin{itemize}
\item[(i)] the Hamiltonian $H=-\frac{1}{2}\Delta+V$ has no bound
states, i.e. $\mathcal{H}_{\text{p.p.}}=\{0\},$ \item[(ii)]
the Hamiltonian $H=-\frac{1}{2}\Delta+V$ has no zero energy
resonances, \item[(iii)] $V$ is a $C^{\infty}$-function on
$\RRR^3$, \item[(iv)] $V$ and its derivatives of all orders are
uniformly bounded in $\bx$: For all multi-indices $\alpha$ there
exist an $M_{\alpha}<\infty$ such that
$|\partial_{\bx}^{\alpha}V(\bx)|<M_{\alpha}$ for all
$\bx\in\RRR^3,$ \item[(v)] there exist positive numbers $\delta$
and $C$ such that
\begin{equation*}
|V(\bx)|\leq C\lxr^{-5-\delta}\text{ for all }\bx\in\RRR^3.
\end{equation*}

\end{itemize}
\ede

\bthe\label{thedasisses} Let $\psi$ be a normalized vector in
$\schwarz$ and suppose that $V$ is in $\mathcal{V}.$ Furthermore,
suppose that the point process
$(\Lambdastar,\mathfrak{F},\mathbb{P})$ satisfies (\ref{expnb}),
(\ref{mu}) and the ergodic assumption (\ref{lawlarge}). Let
$\bko||\be_3$ with $k_0>0$ and suppose that $\bko\notin
C_{\Sigma}.$  Then $\sigemp^\epsilon$ is well defined and
(recalling (\ref{rums}))
\begin{align}\label{dasisses}\sigma_{\text{emp}}^{\epsilon}(\Sigma)=\frac{\ndstar(\dte,R,\Aeps,\Leps,\psieps,\Sigma)}{\dteb}
\overset{\mathbb{P}}{\underset{\epsilon\to 0,R\to\infty,\dteb\to\infty}{\longrightarrow}}
\siglim(\Sigma)=\int\limits_{\Sigma}\siglimdiff(\bom)d\Omega,
\end{align}
where $\siglimdiff(\bom)=16\pi^4
|T(k_0\boldsymbol{\omega},\bko)|^2$ and $\overset{\mathbb{P}}{\longrightarrow}$ denotes convergence in probability. \ethe

\bde\label{defpotstronglight} $V$ is in $\mathcal{V'}$ if
\begin{itemize}
\item[(i)] the Hamiltonian $H=-\frac{1}{2}\Delta+V$ has no bound
states, i.e. $\mathcal{H}_{\text{p.p.}}=\{0\},$ \item[(ii)]
the Hamiltonian $H=-\frac{1}{2}\Delta+V$ has no zero energy
resonances, \item[(iii)] $V$ is in $(V)_5,$ \item[(iv)] $V$ is
$C^{\infty}$ except, perhaps, at a finite number of singularities.
\end{itemize}

\ede

Under these conditions we obtain

\bthe\label{thedasisseslight} Let $\psi$ be a normalized vector in
$C_0^{\infty}(\RRR^3)$ and let $V$ be in $\mathcal{V'}.$
Furthermore, suppose that the point process
$(\Lambdastar,\mathfrak{F},\mathbb{P})$ satisfies (\ref{expnb}),
(\ref{mu}) and the ergodic assumption (\ref{lawlarge}). Let
$\bko||\be_3$ with $k_0>0$ and suppose that $\bko\notin
C_{\Sigma}.$ Then $\sigemp^\epsilon$ is well defined and
(\ref{dasisses}) of Theorem \ref{thedasisses} holds.\ethe

\section{Proof of Theorem 1 and Theorem 2}\label{chformula}

During the proof in this section and in the appendix $0<c<\infty$ will denote a constant
\z{whose value can change during a calculation---even within the same
equation or inequality.}

If either $V\in\mathcal{V}$ and $\psi\in\schwarz$ or $V\in\mathcal{V'}$ and
$\psi\in C_0^\infty$, \z{then (if $\psi$ is normalized) the $\psieps_{\by}$
are ``good'' for all $\by\in\Aeps$ for all $\epsilon$ small enough. That the
$\psieps_{\by}$ satisfy the conditions for the FAST follows from Lemma
\ref{lemmapping} below, and that they satisfy A\ref{condcyc} is easily
seen: For the case $V\in\mathcal{V}$ and $\psi\in\schwarz$ the conclusion
follows from a simple computation, and if $V\in\mathcal{V'}$ and $\psi\in
C_0^\infty$ it suffices to observe that by choosing $\epsilon$ small enough
the wave function $\psieps_{\by}$ has, for all $\by\in\Aeps$, no overlap
with the singularities of the potential.}

\z{$\ndstar$ is thus well defined by (\ref{ndet}), and we can take the
first limit in (\ref{dasisses}) using the following}

\bprop\label{proplargenumber} \z{Suppose that $\psieps_{\by}$ satisfies
A\ref{condcyc} for all $\by\in\Aeps$ and that the potential satisfies A\ref{condpot}. Furthermore, suppose that} the point process
$(\Lambdastar,\mathfrak{F},\mathbb{P})$ satisfies (\ref{expnb}), (\ref{mu})
and the ergodic assumption (\ref{lawlarge}). Then the number of detected
particles $\ndstar(\dte)$ obeys the law of large numbers, i.e. for all
$\delta>0$
\begin{equation}\label{lawlargedet} 
\lim\limits_{\dteb\to\infty}\mathbb{P}\left(\left|\frac{\ndstar(\dte,\Sigma)}{\dteb}-\gamma\right|\geq\delta\right)=0,
\end{equation}
where
\begin{equation}\label{gamma}
\gamma=\int\limits_{\Aeps}\mathbb{E}\left(\nd^{\psi_{\by}^{\epsilon}}(\Sigma)\right)d^2y.
\end{equation}

\eprop

\brem\label{rempoisson} $\gamma=\gamma(\Sigma)$ is in fact the
cross section which would be measured in an experiment. The
remaining limits in (\ref{dasisses}) applied to $\gamma$ yield the
cross section $\siglim(\Sigma)$. If the basic point process is a
Poisson process with $[0,\dte)=\RRR^+$ the times of detection in
$\Sigma$ form a Poisson process with intensity $\gamma.$ Moreover,
in the scattering regime, the detailed detection events, involving
times and directions, form a Poisson process on $\RRR^+\times S^2$
with intensity $\siglimdiff(\bom)$.\erem \bpro{} By the definition
(\ref{ndet}) of $\ndstar$ we have that
\begin{equation}\label{poisson1}
 \ndstar(\dte)(\lambdastar)=\charac_{B(\tau)}^{\star}(\lambdastar)=\sum\limits_{\lambda\in\lambdastar}\charac_{B(\tau)}(\lambda),
\end{equation}
with $B(\tau)$ given by
\begin{equation}\label{poisson2}
 B(\tau)=\{(t,\by,\bq)\in\Lambda|N_{\text{det}}(\tau,\Sigma)(t,\by,\bq)=1\}.
\end{equation}
It thus follows from (\ref{expnb}) and (\ref{mu}) that
\begin{align}\label{poisson3}
 \mathbb{E}\left(\ndstar(\dte)\right)=&\mu(B(\tau))=\int\charac_{[0,\tau)}(t)\nd^{\psiyeps}(\Sigma)(\bq)d\mu=\dteb\int\limits_{\Aeps}\mathbb{E}\left(\nd^{\psiyeps}(\Sigma)\right)d^2y=\dteb\gamma.
\end{align}
The proposition follows from the ergodicity assumption
(\ref{lawlarge}). \epro

It is not easy to calculate the expectation value $\gamma$ (cf.
(\ref{gamma})) directly. \z{However, as we shall show below, using the FAST
we can approximate (\ref{gamma}) by
\begin{equation}\label{tsche2}
  \int\limits_{\Aeps}\mathbb{E}\left(\nsi^{\psiyeps}(R\Sigma)\right)d^2y,
\end{equation}
where the integrand of (\ref{tsche2}) is given by an integral over the flux
(cf. (\ref{expnsig})), an expression that we can more easily handle.
We} will show in Lemma \ref{lemuniform} below that
$\mathbb{E}\left(\nsi^{\psiyeps}(R\Sigma)\right)$ is absolutely integrable
over $\Aeps.$ 

\z{We introduce} now a class of scattering states $\g$ for which
we can show that the corresponding asymptotes are in the set $\ghut$,
i.e. that they satisfy the FAST.

\bde\label{defgp} A function $f:\RRR^3\to\CCC$ is in $\gprime$ if
\footnote{$C^8(H):=\bigcap\limits_{n=1}^{8}\dom(H^n)$}
\begin{equation*}f\in\mathcal{H}_{\text{a.c.}}(H)\cap C^8(H),\end{equation*}
\begin{equation*}\lxr^2 H^nf(\bx)\in\ltr,\;n\in\{0,1,2,...,8\},\end{equation*} \begin{equation*}\lxr^4 H^nf(\bx)\in\ltr,\;n\in\{0,1,2,3\}.\end{equation*} Then $\g:=\bigcup\limits_{t\in\RRR}e^{-iHt}\gprime.$\ede

We state now the important lemma \z{that ensures that the $\psiyeps$
satisfy} the FAST.

\blem\label{lemmapping} Suppose $V\in(V)_4$ and that zero is
neither a resonance nor an eigenvalue of $H$. Then
\begin{equation*}
\psi(\bx)\in\g\Rightarrow\widehat{\psi}_{\text{out}}(\bk)=\mathcal{F}\left(\Omega_+^{-1}\psi\right)(\bk)
\in\ghut.
\end{equation*}

\elem
\z{The proof} is adapted from \cite{duerr:03} and can be found in the
appendix. 

\brem For other mapping properties between $\psi$ and $\pout$,
which are not applicable in our case, see \cite{yajima:95}.\erem

For $\psi\in\mathcal{S}$ and $V\in\mathcal{V}$ or $\psi\in
C_0^\infty(\RRR^3)$ and $V\in\mathcal{V'}$ we have that $\psieps_{\by}\in
C^{\infty}(H)$ for all $\by\in\Aeps$ and $\epsilon$ small enough. By (i) in
the definition of $\mathcal{V}$ or $\mathcal{V'}$ (Definition
\ref{defpotstrong} or \ref{defpotstronglight}) there are no bound
states. Hence $\psieps_{\by}\in\mathcal{H}_{\text{a.c.}}(H)\cap C^8(H)$,
\z{and one easily sees that $\psieps_{\by}\in\g$.} Thus \z{ by Lemma
\ref{lemmapping} and Proposition \ref{propflux} the} $\psiyeps$ satisfy the
FAST for all $\by\in\Aeps$ and $\epsilon$ small enough. 

\z{We now} show that $\mathbb{E}\left(\nsi^{\psiyeps}(R\Sigma)\right)$ is
absolutely integrable over ${\Aeps}.$

\blem\label{lemuniform} \z{Suppose that $\psi\in\mathcal{S}$ and
$V\in\mathcal{V}$ or that} $\psi\in
C_0^\infty(\RRR^3)$ and $V\in\mathcal{V'}$. Then there
exist $M$ and $R_0>0$ such that for $\epsilon$ small enough
\begin{equation}\label{uniformy}
\int\limits_0^{\infty}\int\limits_{ RS^2}
|\bj^{\psi_{\by,t}^{\epsilon}}(\bx)\cdot d\bsig| dt<M,\;\forall\by\in
{\Aeps},\forall R>R_0.
\end{equation}

\elem
For the proof see the Appendix. From now on we assume that
$R>R_0.$

By Lemma \ref{lemmapping}, Proposition \ref{propflux}, Proposition
\ref{propexpn} and Lemma \ref{lemuniform} we see that
(\ref{tsche2}) is a well defined expression. Moreover, by
(\ref{uni1}) the difference between
$\mathbb{E}\left(\nd^{\psiyeps}(R,\Sigma)\right)$ and
$\mathbb{E}\left(\nsi^{\psiyeps}(R\Sigma)\right)$ vanishes in the
limit $R\to\infty$, and using Lemma \ref{lemuniform} we easily see
by the dominated convergence theorem that the same conclusion
holds for the integrals themselves. Thus, by Proposition
\ref{proplargenumber}, the limit $\siglim(\Sigma)$ in Theorem
\ref{thedasisses} is given by
\begin{align}\label{familiar0}
 \siglim(\Sigma)=&\lim\limits_{\epsilon\to
0}\lim\limits_{R\to\infty}\gamma=\lim\limits_{\epsilon\to
0}\lim\limits_{R\to\infty}\int\limits_{\Aeps}\mathbb{E}\left(\nd^{\psiyeps}(R,\Sigma)\right)d^2y=\lim\limits_{\epsilon\to
0}\int\limits_{\Aeps}\lim\limits_{R\to\infty}\mathbb{E}\left(N_{\text{sig}}^{\psiyeps}(R\Sigma)\right)d^2y\notag\\=&\lim\limits_{\epsilon\to
0}\int\limits_{\Aeps}\lim\limits_{R\to\infty}\int\limits_0^\infty\int\limits_{R\Sigma}\bj^{\psi_{\by,t}^{\epsilon}}(\bx)\cdot
d\bsig dtd^2y.
\end{align}
Using Lemma \ref{lemmapping} and Proposition \ref{propflux} we get
instead of (\ref{familiar0})
\begin{align}\label{familiar}
 \siglim(\Sigma)&=\lim\limits_{\epsilon\to 0}\int\limits_{C_\Sigma}\int\limits_{\Aeps}
|\widehat{\Omega_+^{-1}\psi^{\epsilon}_{\by}}(\bk)|^2d^2yd^3k=\lim\limits_{\epsilon\to 0}\int\limits_{C_\Sigma} \int\limits_{\Aeps}
|\widehat{S\Omega_-^{-1}\psi^{\epsilon}_{\by}}(\bk)|^2d^2yd^3k.
\end{align}
The formula for $S=T+I$ is given by (\ref{formulat}) and
(\ref{tmatrix}). To exploit this formula we write instead of
(\ref{familiar}):
\begin{equation}\label{schwarz1}
 \siglim(\Sigma)=\lim\limits_{\epsilon\to 0}\int\limits_{C_\Sigma}
\int\limits_{\Aeps}
|\mathcal{F}\left(S(\Omega_-^{-1}\psiyeps-\psiyeps)+T\psiyeps+\psiyeps\right)(\bk)|^2d^2yd^3k.\end{equation}
By \z{the} triangle equality we see that (\ref{schwarz1}) yields
\begin{equation}\label{schwarz2}
 \siglim(\Sigma)=\lim\limits_{\epsilon\to
0}\int\limits_{C_\Sigma}\int\limits_{\Aeps}  
|\mathcal{F}(T\psiyeps)(\bk)|^2d^2yd^3k,
\end{equation}
provided
\begin{equation}\label{schwarz3}
\lim\limits_{\epsilon\to 0}\int\limits_{\Aeps}
\|\Omega_-^{-1}\psiyeps-\psiyeps\|^2d^2y=0
\end{equation}
and
\begin{align}\label{tail0}
 \lim\limits_{\epsilon\to 0}
\int\limits_{C_\Sigma}\int\limits_{\Aeps}
|\widehat{\psi_{\by}^{\epsilon}}(\bk)|^2d^2yd^3k=0.
\end{align}

\brem In \cite{duerr:00} the ``sufficient condition" for
proceeding from (\ref{familiar}) to (\ref{schwarz2}) was
insufficient.\erem

We will establish now (\ref{schwarz3}) and (\ref{tail0}). We start
with (\ref{tail0}). Suppose that $\Sigma$ is such that $\bko\notin
C_{\Sigma}.$ With (\ref{psiepsfou}) we have then that
\begin{align}\label{tail1}
  \int\limits_{C_\Sigma}\int\limits_{\Aeps}
|\widehat{\psi_{\by}^{\epsilon}}(\bk)|^2d^2yd^3k=&\epsilon^{-3}\int\limits_{C_\Sigma}\int\limits_{\Aeps}
|\widehat{\psi}\left(\frac{\bk-\bko}{\epsilon}\right)|^2d^2yd^3k=\int\limits_{\frac{1}{\epsilon}(C_\Sigma-\bko)}\int\limits_{\Aeps}
|\widehat{\psi}\left(\bk\right)|^2d^2yd^3k.
\end{align}
Since $\bko\notin C_{\Sigma}$ there exists a $\delta>0$ such that
\begin{align}\label{tail2}
  |\bk-\bko|>\delta\text{ for all }\bk\in C_{\Sigma}.
\end{align}
Using that $\widehat{\psi}\in\schwarz$ (we will use that
$|\widehat{\psi}|\leq \con k^{-(d+2)}$), the last integral in
(\ref{tail1}) can be estimated by
\begin{align}\label{tail3} 
\int\limits_{\frac{1}{\epsilon}(C_\Sigma-\bko)}\int\limits_{\Aeps}
|\widehat{\psi}\left(\bk\right)|^2d^2yd^3k\leq&\int\limits_{k>\frac{\delta}{\epsilon}}\int\limits_{\Aeps}
|\widehat{\psi}\left(\bk\right)|^2d^2yd^3k\leq\frac{\con}{\epsilon^{2d}}\int\limits_{k>\frac{\delta}{\epsilon}}
\frac{1}{k^{2d+4}}d^3k\leq \con\epsilon,
\end{align}
from which (\ref{tail0}) follows.

Since $\Omega_-$ is a partial isometry, (\ref{schwarz3}) is
equivalent to
\begin{equation}\label{schwarz4}
\lim\limits_{\epsilon\to 0}\int\limits_{\Aeps}
\|\Omega_-\psiyeps-\psiyeps\|^2d^2y=0,
\end{equation}
which is the content of the following

\blem\label{lemint} Let zero be neither an eigenvalue nor a
resonance of $H$ and suppose that $V\in (V)_5$. Let $\psi \in \schwarz$ and let $k_0>0$. Then
\begin{equation}\label{theb}
\lim\limits_{\epsilon\to 0}\int\limits_{\Aeps}
\|\Omega_-\psiyeps-\psiyeps\|^2d^2y=0.
\end{equation}

\elem

\brem\label{remint} Under the additional condition that
$\supp\widehat{\psi}\subset P_{\be_3}^{\alpha}$ for some
$\alpha\in(0,\frac{\pi}{2})$, where
$P_{\be_3}^{\alpha}:=\{\bk\in\RRR^3:\bk\cdot\be_3>k\cos\alpha\}$,
$0<\alpha<\frac{\pi}{2}$ (\z{this  is a convenient}
condition, see e.g. \cite{amrein:77}, Lemma 7.17), one can prove
in a manner similar to the way we prove Lemma \ref{lemint} that
the following holds:
\begin{equation}\label{the}
\lim\limits_{L\to\infty}\int\limits_{\GammaL}
\|\Omega_-\psi_{\by}-\psi_{\by}\|^2d^2y=0.
\end{equation}
It is well known that the integrand in (\ref{the}) tends to zero
for large $y$ (see e.g. \cite{amrein:77}, Corollary 8.17,
\cite{simon3:79}, Theorem XI.33, and \cite{teufel1:99}, Theorem
2.20).\erem 
\bpro{ of Lemma \ref{lemint}} We have that
\begin{equation}\label{norma}
\|\Omega_-\psiyeps-\psiyeps\|^2=1-(\psiyeps,\Omega_-\psiyeps)+c.c.
\end{equation}
Since $\Omega_-\psi=\mathcal{F}_-^{-1}\widehat{\psi}(\bk)$ for any
$\psi\in \ltr$ (Proposition \ref{propexpansion}, (iii)) we obtain
for the r.h.s. of (\ref{norma}):
\begin{equation}\label{f1f2a}
1-\int(\psiyeps)^{\ast}(\bx)(2\pi)^{-3/2}\int\widehat{\psi_{\by}^\epsilon}(\bk)\varphi_-(\bx,
\bk)d^3kd^3x+c.c.
\end{equation}
Writing
\begin{equation}\label{etaphi}
\varphi_-(\bx,\bk)=e^{i\bk\cdot\bx}-\eta_-(\bx,\bk)
\end{equation}
and since $\|\psi_{\by}^{\epsilon}\|^2=1$, we then find that
\begin{align}\label{expra}
\|\Omega_-\psiyeps-\psiyeps\|^2=&\int(\psiyeps)^{\ast}(\bx)(2\pi)^{-3/2}\int\widehat{\psi_{\by}^\epsilon}(\bk)
\eta_-(\bx,\bk)d^3kd^3x+c.c.
\end{align}

We shall divide the $\bk$-integration into two parts with the help
of smooth \z{($C^\infty$) mollifiers $0\leq f_1(\bk)\leq 1$ and $0\leq
f_2(\bk)\leq 1$ satisfying}
\begin{align}\label{deff1f2}
f_1(\bk)=&
\begin{cases}
1,\text{ for }|\bk-\bko|<\frac{k_0}{3},\\0,\text{ for
}|\bk-\bko|\geq\frac{k_0}{2},
\end{cases}\notag\\f_2(\bk):=&1-f_1(\bk).
\end{align}
Using (\ref{deff1f2}) we obtain for (\ref{expra})
\begin{align}\label{f1f2b}
\|\Omega_-\psiyeps-\psiyeps\|^2=&\int(\psiyeps)^{\ast}(\bx)(2\pi)^{-3/2}
\int\widehat{\psi_{\by}^\epsilon}(\bk)(f_1+f_2)(\bk)\eta_-(\bx,
\bk)d^3kd^3x+c.c.\notag\\=&\int(\psiyeps)^{\ast}(\bx)(2\pi)^{-3/2}
\int\widehat{\psi_{\by}^\epsilon}(\bk)f_1(\bk)\eta_-(\bx,
\bk)d^3kd^3x\notag\\&+\int(\psiyeps)^{\ast}(\bx)(2\pi)^{-3/2}
\int\widehat{\psi_{\by}^\epsilon}(\bk)f_2(\bk)\eta_-(\bx,
\bk)d^3kd^3x+c.c.=:I_1+I_2+c.c.\notag\\\leq&2|I_1|+2|I_2|.
\end{align}

Observing that $\psi\in\schwarz$ we estimate $|I_2|$ by using that
for any $n>0$ $|\widehat{\psi}(\bk)|\leq\frac{\con}{k^n}$ and that
$|\eta_-(\bx,\bk)|\leq 1+|\varphi_-(\bx,\bk)|\leq \con$ (Proposition
\ref{proplippschw} (ii)) as well as (\ref{psieps}),
(\ref{psiepsfou}) and (\ref{deff1f2}):
\begin{align}\label{expraab}
|I_2|\leq&\frac{\con}{\epsilon^3}\int|\psi(\bx-\by)|(2\pi)^{-3/2}\int\limits_{|\bk-\bko|\geq\frac{k_0}{3}}\left|\widehat{\psi}
\left(\frac{\bk-\bko}{\epsilon}\right)\right|d^3kd^3x\notag\\\leq&\frac{\con}{\epsilon^3}\int\limits_{|\bk|\geq\frac{k_0}{3}}
\left|\widehat{\psi}\left(\frac{\bk}{\epsilon}\right)\right|d^3k\leq
c\epsilon^{n-3}\int\limits_{|\bk|\geq\frac{k_0}{3}}\frac{1}{k^n}d^3k=\con\epsilon^{n-3},\end{align}
if $n\geq 4.$

Lemma \ref{lemint} concerns the integration of $I_1$ and $I_2$
over $\Aeps$. With (\ref{expraab}) we obtain that
\begin{align}\label{expraac}
\int\limits_{\Aeps}|I_2|d^2y\leq&\con\epsilon^{n-3-2d},
\end{align}
which tends to zero if we choose $n$ large
enough. We are left with showing that
\begin{align}\label{schwarz3a}
\lim\limits_{\epsilon\to 0}\int\limits_{\Aeps}|I_1|d^2y=0,
\end{align}
and for this it suffices to prove that
\begin{align}\label{schwarz3b} \lim\limits_{\epsilon\to
0}\int\limits_{Y_{\Leps}}|I_1|d^2y=0.
\end{align}

Recalling the Lippmann-Schwinger equation (\ref{lippmann}), i.e.
that
\begin{equation*}
\eta_-(\bx,\bk)\hspace{-0,2pt}=\hspace{-0,2pt}\frac{1}{2\pi}\int\gre
V(\bxp)\varphi_-(\bxp,\bk),
\end{equation*}
we find that
\begin{equation}\label{exprbb}
I_1=\frac{1}{(2\pi)^{\frac{5}{2}}}\hspace{-1.5pt}\int\hspace{-1.5pt}
(\psieps_{\by})^{\ast}(\bx)\hspace{-1.5pt}
\int\hspace{-1.5pt}\widehat{\psieps_{\by}}(\bk)f_1(\bk)\hspace{-1.5pt} \int\gre
V(\bxp)\varphi_-(\bxp,\bk)d^3x'd^3kd^3x.
\end{equation}
Since the integrand in (\ref{exprbb}) is absolutely integrable
over $\bx,\bxp,\bk$ (because $\psi\in\schwarz$, $V\in (V)_5$; cf.
Lemma \ref{proplippschw}, (ii)) we are free to interchange these
integrations and more generally change integration variables as
convenient. Using
$(\psieps_{\by})^{\ast}(\bx)=(\psieps)^{\ast}(\bx-\by),
\widehat{\psieps_{\by}}(\bk)=\widehat{\psieps}(\bk)e^{-i\bk\cdot\by}$
we obtain \z{that}
\begin{equation}\label{cond3}
\begin{split}
I_1=\frac{1}{(2\pi)^{\frac{5}{2}}}
\int\limits_{\RRR^3}(\psieps)^{\ast}(\bx-\by)\int\limits_{\RRR^3}\widehat{\psieps}(\bk)f_1(\bk)
\int\limits_{\RRR^3}\frac{e^{ik|\bx-\bxp|-i\bk\cdot\by}}{|\bx-\bxp|}
V(\bxp)\varphi_-(\bxp,\bk)d^3x'd^3kd^3x.
\end{split}
\end{equation}
Making the change of variables $\bx\to\bx-\by$ and using
$\by=(y_1,y_2,-\Leps)$ we obtain
\begin{equation}\label{cond4}
\begin{split}
I_1=\frac{1}{(2\pi)^{\frac{5}{2}}}
\int\limits_{\RRR^3}(\psieps)^{\ast}(\bx)\int\limits_{\RRR^3}\widehat{\psieps}(\bk)f_1(\bk)
\int\limits_{\RRR^3}\frac{e^{ik|\by+\bx-\bxp|-ik_1y_1-ik_2y_2+ik_3\Leps}}{|\by+\bx-\bxp|}
V(\bxp)\cdot \\ \cdot\varphi_-(\bxp,\bk)d^3x'd^3kd^3x.
\end{split}
\end{equation}
Introducing as shorthand notation (no change of variables)
$\tilde{\by}=\by+\bx-\bxp$, $\ba:=\bx-\bxp$, $b_3:=-\Leps+a_3$ and
letting $(r,\theta)$ be the polar coordinates for
$(\tilde{y}_1,\tilde{y}_2)$, with $\be_r$ the corresponding radial
unit vector $(\bot\be_3)$, this becomes
\begin{align}\label{polar1}
I_1=&\frac{1}{(2\pi)^{\frac{5}{2}}}
\int\limits_{\RRR^3}(\psieps)^{\ast}(\bx)\int\limits_{\RRR^3}\widehat{\psieps}(\bk)f_1(\bk)
\int\limits_{\RRR^3}\frac{e^{ik\sqrt{\tilde{y}_1^2+\tilde{y}_2^2+(-\Leps+a_3)^2}-ik_1\tilde{y}_1-ik_2\tilde{y}_2+ik_3\Leps} }{|\tilde{\by}|}
e^{ik_1a_1+ik_2a_2}\cdot\notag\\ &\hspace{8,65cm}\cdot
V(\bxp)\varphi_-(\bxp,\bk)d^3x'd^3kd^3x\notag\\=&\frac{1}{(2\pi)^{\frac{5}{2}}}
\int\limits_{\RRR^3}(\psieps)^{\ast}(\bx)\int\limits_{\RRR^3}\widehat{\psieps}(\bk)f_1(\bk)\int\limits_{\RRR^3}
\frac{e^{ik\sqrt{r^2+b_3^2}-ik\sin \vartheta\: r\cos \beta+ik\cos\vartheta
\Leps}}{\sqrt{r^2+b_3^2}} e^{ik_1a_1+ik_2a_2}\cdot\notag\\
&\hspace{7,6cm}\cdot V(\bxp)\varphi_-(\bxp,\bk)d^3x'd^3kd^3x,
\end{align}
with $k\sin\vartheta=|\bk_p| =\sqrt{k_1^2+k_2^2}$, $k_3=k\cos\vartheta$,
where $\vartheta$ $(0\leq\vartheta\leq\pi)$ is the angle between $\bk$ and
$\be_3$ and $\beta$ is the angle between $\bk_p=(k_1,k_2,0)$ and $\be_r.$
\z{Moreover, there is an angle $0<\alpha<\frac{\pi}{2}$ such that
\begin{equation}\label{support}
  \vartheta\leq\alpha,\text{ i.e. }\cos\alpha\leq\cos\vartheta\leq
1,\;0\leq\sin\vartheta\leq\sin\alpha,\;0<\alpha<\frac{\pi}{2}
\end{equation}
for all $\bk$'s in the support of $f_1$ (cf. (\ref{deff1f2})).}

We introduce now spherical coordinates $(k,\bom)$ for $\bk$ as
integration variables and do first the integration over $k$ (note
that $\beta$ is not $k$-dependent). Since
$\widehat{\psieps}\in\schwarz$, $f_1$ is smooth and $\frac{\partial}{\partial k}\varphi_-(\bxp,\bk)$ is
uniformly bounded in $k$ (Proposition \ref{proplippschw} (iv)), we
can do two integration by parts \z{with respect to}  $k$ and obtain that
\begin{align}\label{polar3}
I_1=&-\frac{1}{(2\pi)^{\frac{5}{2}}}
\int\limits_{\RRR^3}(\psieps)^{\ast}(\bx) \int\limits_{\RRR^3}
V(\bxp)\int\limits_{S^2}
\int\limits_0^{\infty}\frac{\partial^2}{\partial k^2}
\left(\widehat{\psieps}(\bk)f_1(\bk)\varphi_-(\bxp,\bk)e^{ik_1a_1+ik_2a_2}k^2\right)\cdot \notag
\\ &\hspace{7,55cm}\cdot\frac{e^{ik\lambda}}{\sqrt{r^2+b_3^2}\lambda^2} dkd\Omega d^3x'd^3x,
\end{align}
where
\begin{equation}\label{lam}
 \lambda:=r(\sqrt{1+\frac{b_3^2}{r^2}}-\sin
\vartheta\cos\beta)+\cos\vartheta \Leps.
\end{equation}

To estimate the derivatives of \z{the functions
$f_1(\bk)\varphi_-(\bxp,\bk)$ we use Proposition \ref{proplippschw}, (iv)
and the smoothness of $f_1(\bk)$.}  We introduce a multi-index \z{notation}
\begin{equation*}
  i:=(i_1,i_2,i_3,i_4),\; i_m\in\NNN_0,\; |i|:=i_1+i_2+i_3+i_4,
\;j:=(j_1,j_2,j_3)\text{ analogously}.
\end{equation*}
With $k_l=\kappa_l k,\;\kappa_l\in[-1,1],\: l=1,2$ we obtain that
\begin{align}\label{derivatives}
\left|\frac{\partial^2}{\partial
k^2}(f_1(\bk)\varphi_-(\bxp,\bk)\widehat{\psieps}(\bk)k^2e^{ik_1a_1+ik_2a_2})\right|\notag\\&\hspace{-4,5cm}\leq2\sum\limits_{|i|=2}\left|\frac{\partial^{i_1}}{\partial
k^{i_1}}\left(f_1(\bk)\varphi_-(\bxp,\bk)\right)\right|\left|\frac{\partial^{i_2}}{\partial
k^{i_2}}(\widehat{\psieps}(\bk)k^2)\right|\left|\frac{\partial^{i_3}}{\partial
k^{i_3}}(e^{i\kappa_1 ka_1})\right|\left|\frac{\partial^{i_4}}{\partial
k^{i_4}}(e^{i\kappa_2 ka_2})\right|\notag\\ &\hspace{-4,5cm}\leq
c\sum\limits_{|i|=2}(1+x')^{i_1}\left|\frac{\partial^{i_2}}{\partial
k^{i_2}}(\widehat{\psieps}(\bk)k^2)\right|\left|\kappa_1
a_1|^{i_3}|\kappa_2 a_2\right|^{i_4}\notag\\ &\hspace{-4,5cm}\leq
c\sum\limits_{|i|=2}(1+x')^{i_1}\left|\frac{\partial^{i_2}}{\partial
k^{i_2}}(\widehat{\psieps}(\bk)k^2)\right|a^{i_3}a^{i_4}\notag\\
&\hspace{-4,5cm}\leq
c\sum\limits_{|i|=2}(1+x')^{i_1}\left|\frac{\partial^{i_2}}{\partial
k^{i_2}}(\widehat{\psieps}(\bk)k^2)\right|\left|\bx-\bxp\right|^{i_3+i_4}\notag\\
&\hspace{-4,5cm}\leq
c\sum\limits_{|j|=2}(1+x')^{j_1}\left|\frac{\partial^{j_2}}{\partial
k^{j_2}}(\widehat{\psieps}(\bk)k^2)\right|\left|\bx-\bxp\right|^{j_3}.
\end{align}

With (\ref{support}) \z{we may assume that} $\lambda$ in
(\ref{lam}) is bounded below,
\begin{equation}\label{eta}
\lambda\geq
r(1-\sin\alpha)+\Leps\cos\alpha\geq\lambda_{\text{min}}:=\eta(r+\Leps),
\end{equation}
with $\eta:=\min((1-\sin\alpha),\cos\alpha)>0.$ \z{Using (\ref{eta}) and
(\ref{derivatives}) in (\ref{polar3})} we obtain \z{that}
\begin{align}\label{polar4}
 M:=\int\limits_{Y_{\Leps}}|I_1|d^2y&
\leq\con\sum\limits_{|j|=2}
\int\limits_{\RRR^2} \int\limits_{\RRR^3}|\psieps(\bx)|
\int\limits_{\RRR^3} |V(\bxp)|\int\limits_{S^2}
\int\limits_0^{\infty}\frac{1}{\sqrt{r^2+b_3^2}\lambda_{\text{min}}^2}\left|\partial_k^{j_2}\left(\widehat{\psieps}(\bk)k^2\right)\right|\notag
\\ &\hspace{4,8cm}|\bx-\bxp|^{j_3}(1+x')^{j_1}
dkd\Omega d^3x'd^3xd^2y.
\end{align}
Since the integrand of the \z{right hand side} of (\ref{polar4}) is positive,
we may perform the change of integration
variables $(y_1,y_2)\to(\tilde{y}_1,\tilde{y}_2)\to(r,\theta),$ as
well as freely interchange the order of integrations. With
(\ref{eta}) we then obtain \z{that}
\begin{align}\label{polar41}
  M & \leq\con\sum\limits_{|j|=2}  \int\limits_{\RRR^3}|\psieps(\bx)|
\int\limits_{\RRR^3} |V(\bxp)|\int\limits_{S^2}
\int\limits_0^{\infty}\int\limits_0^{\infty}\int\limits_0^{2\pi}
\frac{1}{\sqrt{r^2+b_3^2}\lambda_{\text{min}}^2}\left|\partial_k^{j_2}\left(\widehat{\psieps}(\bk)k^2\right)\right|\notag \\
&\hspace{5cm}|\bx-\bxp|^{j_3}(1+x')^{j_1}
rd\theta dr dkd\Omega d^3x'd^3x\notag \\&\leq \con\sum\limits_{|j|=2}
\int\limits_{\RRR^3}|\psieps(\bx)| \int\limits_{\RRR^3}
|V(\bxp)|\int\limits_{S^2}
\int\limits_0^{\infty}\int\limits_0^{\infty}\frac{1}{\eta^2(r+\Leps)^2}\left|\partial_k^{j_2}\left(\widehat{\psieps}(\bk)k^2\right)\right|\notag
\\
&\hspace{4,75cm}|\bx-\bxp|^{j_3}(1+x')^{j_1}
drdkd\Omega d^3x'd^3x\notag\\&=\frac\con{\eta^2 \Leps}\sum\limits_{|j|=2}
\int\limits_{\RRR^3}|\psieps(\bx)| \int\limits_{\RRR^3}
|V(\bxp)|\int\limits_{S^2}
\int\limits_0^{\infty}\left|\partial_k^{j_2}\left(\widehat{\psieps}(\bk)k^2\right)\right||\bx-\bxp|^{j_3}(1+x')^{j_1} dkd\Omega d^3x'd^3x.
\end{align}
Using that $|\bx-\bxp|^{j_3}\leq 2(x^{j_3}+x'^{j_3})$ for
$j_3=1,2$ we obtain that
\begin{align}\label{polar5}
  M & \leq\frac\con{\Leps} \sum\limits_{|j|=2}\int\limits_{\RRR^3}|\psieps(\bx)|(1+x)^{j_3}\int\limits_{\RRR^3}\left|\partial_k^{j_2}\left(\widehat{\psieps}(\bk)k^2\right)\right|\int\limits_{\RRR^3} |V(\bxp)|(1+x')^{j_1+j_3}d^3x' dkd\Omega d^3x.
\end{align}
Since $V\in(V)_5$ (so that $V\in\ltr$ and $|V(\bx)|\leq C x^{-5-\delta},\;\delta>0$, for $x> R_0$)  and $j_1+j_3\leq 2$ the $\bxp$ integration is finite and we obtain (by dividing the integration region for $\bxp$ into two parts, $x'>R_0$ and
$x'\leq R_0$)
\begin{align}\label{polar6}
  M & \leq\frac\con{\Leps}\sum\limits_{j_2+j_3\leq 2} \int\limits_{\RRR^3}|\psieps(\bx)|(1+x)^{j_3}\int\limits_{\RRR^3}\left|\partial_k^{j_2}\left(\widehat{\psieps}(\bk)k^2\right)\right|dkd\Omega d^3x.
\end{align}
Using (\ref{psieps}), (\ref{psiepsfou}) and that $\psi\in\schwarz$
one finds by simple calculation that
\begin{align}\label{polar7}
\int\limits_{\RRR^3}|\psieps(\bx)|x^{j_3}d^3x\leq& \frac\con{\epsilon^{\frac{3}{2}}}\frac{1}{\epsilon^{j_3}}
\end{align}
\z{and}
\begin{align}\label{polar7a}
\int\limits_{\RRR^3}\left|\partial_k^{j_2}\left(\widehat{\psieps}(\bk)k^2\right)\right|dkd\Omega\leq& \con\epsilon^{\frac{3}{2}}\frac{1}{\epsilon^{j_2}}.
\end{align}
Since $j_2+j_3\leq 2$ we see with (\ref{polar7}), (\ref{polar7a}) and (\ref{leps})
that for $M$ in (\ref{polar6}) we have for small $\epsilon$ the bound 
\begin{align}\label{polar8}
M\leq\frac\con{\Leps\epsilon^2}=\con\epsilon^{l-2}.
\end{align}
\z{Since} $l>2,$ this completes the proof of (\ref{theb}). \epro

We can now proceed with the evaluation of (\ref{schwarz2}). With (\ref{formulat}) we obtain for (\ref{schwarz2})
\begin{align}\label{familiar2}
\siglim(\Sigma)=&\lim\limits_{\epsilon\to 0}
\int\limits_{C_\Sigma}   \int\limits_{\Aeps}  
|\widehat{T\psi_{\by}^{\epsilon}}(\bk)|^2d^2yd^3k\notag\\=&  
\lim\limits_{\epsilon\to 0} 4\pi^2\int\limits_{C_\Sigma} \int\limits_{\Aeps}
\left|\hspace{3pt}\int\limits_{k'=k}e^{-i\bkp\cdot\by}T(\bk,\bkp)
\widehat{\psi^{\epsilon}}(\bkp)k'd\Omega(\bkp)\right|^2d^2yd^3k\notag\\=&
  \lim\limits_{\epsilon\to 0} 4\pi^2\int\limits_{C_\Sigma} \int\limits_{y_p<\Reps}
\left|\hspace{3pt}\int\limits_{k'=k}e^{-i(k_1'y_1+k_2'y_2-k_3'\Leps)}T(\bk,\bkp)
\widehat{\psieps}(\bkp)k'd\Omega(\bkp)\right|^2dy_1dy_2d^3k,
\end{align}
where $\byp:=(y_1,y_2).$ We \z{insert} again the identity $f_1+f_2\equiv 1$ and obtain for $\siglim(\Sigma)$
\begin{align}\label{familiar21}
\lim\limits_{\epsilon\to 0} 4\pi^2\int\limits_{C_\Sigma}
\int\limits_{y_p<\Reps}
\left|\hspace{3pt}\int\limits_{k'=k}e^{-i(k_1'y_1+k_2'y_2-k_3'\Leps)}T(\bk,\bkp)
\widehat{\psieps}(\bkp)(f_1(\bkp)+f_2(\bkp))k'd\Omega(\bkp)\right|^2dy_1dy_2d^3k.
\end{align}
Multiplying out we get four terms. The main term is
\begin{align}\label{familiar22}
\lim\limits_{\epsilon\to 0} 4\pi^2\int\limits_{C_\Sigma} \int\limits_{y_p<\Reps}
\left|\hspace{3pt}\int\limits_{k'=k}e^{-i(k_1'y_1+k_2'y_2-k_3'\Leps)}T(\bk,\bkp)
\widehat{\psieps}(\bkp)f_1(\bkp)k'd\Omega(\bkp)\right|^2dy_1dy_2d^3k.
\end{align}

Before we evaluate (\ref{familiar22}) we show that the three other
terms are zero. Noting that $T(\bk,\bkp)$ is bounded (Corollary \ref{cort})
and that $\psi\in\schwarz$ we obtain that
\begin{align}\label{remaina}
\left|\hspace{3pt}\int\limits_{k'=k}e^{-i(k_1'y_1+k_2'y_2-k_3'\Leps)}T(\bk,\bkp)
\widehat{\psieps}(\bkp)f_i(\bkp)k'd\Omega(\bkp)\right|\leq&
\frac{\con}{\epsilon^{\frac{3}{2}}}
k,
\;i=1,2,\notag\\\left|\hspace{3pt}\int\limits_{k'=k}e^{-i(k_1'y_1+k_2'y_2-k_3'\Leps)}T(\bk,\bkp)
\widehat{\psieps}(\bkp)f_2(\bkp)k'd\Omega(\bkp)\right|\leq&
\frac{\con}{\epsilon^{\frac{3}{2}}}k\int\limits_{k'=k}\left|\widehat{\psi}\left(\frac{\bkp-\bko}{\epsilon}\right)\right|
f_2(\bkp)d\Omega(\bkp).
\end{align}
Using (\ref{remaina}), the difference between (\ref{familiar22}) and
(\ref{familiar21}) is no greater than
\begin{align}\label{remainb}
\frac{\con}{\epsilon^3}\int\limits_{C_\Sigma}\int\limits_{y_p<\Reps}\int\limits_{k'=k}\left|
\widehat{\psi}\left(\frac{\bkp-\bko}{\epsilon}\right)\right|f_2(\bkp)k'^2d\Omega(\bkp)d^2yd^3k\leq&
\frac{\con}{\epsilon^{3+2d}}\int\limits_{\RRR^3}\left|\widehat{\psi}\left(\frac{\bkp-\bko}{\epsilon}\right)\right|
f_2(\bkp)k'^2d^3k'\notag\\\leq&\frac{\con}{\epsilon^{3+2d}}\int\limits_{|\bkp-\bko|\geq\frac{k_0}{3}}
\left|\widehat{\psi}\left(\frac{\bkp-\bko}{\epsilon}\right)\right|k'^2d^3k'.
\end{align}
Using that $|\widehat{\psi}(\bk)|\leq\frac{\con}{k^n}$ \z{for} any $6\leq n\in\NNN$, we see
that the right-hand side in (\ref{remainb}) is bounded by
$\con\epsilon^{n-3-2d}$\z{, which} tends to zero for sufficiently large
$n$. Thus the three other terms are zero.

Since, as we shall show,
\begin{align}\label{familiar24}
\lim\limits_{\epsilon\to 0} 4\pi^2\int\limits_{C_\Sigma} \int\limits_{y_p\geq\Reps}
\left|\hspace{3pt}\int\limits_{k'=k}e^{-i(k_1'y_1+k_2'y_2-k_3'\Leps)}T(\bk,\bkp)
\widehat{\psieps}(\bkp)f_1(\bkp)k'd\Omega(\bkp)\right|^2dy_1dy_2d^3k=0,
\end{align}
we may extend the $\by$-integration in (\ref{familiar22}) to all of $\RRR^2$, so that
\begin{align}\label{familiar23}
\siglim(\Sigma)=\lim\limits_{\epsilon\to 0} 4\pi^2\int\limits_{C_\Sigma} \int\limits_{\RRR^2}
\left|\hspace{3pt}\int\limits_{k'=k}e^{-i(k_1'y_1+k_2'y_2-k_3'\Leps)}T(\bk,\bkp)
\widehat{\psieps}(\bkp)f_1(\bkp)k'd\Omega(\bkp)\right|^2dy_1dy_2d^3k.
\end{align}

Before \z{establishing} (\ref{familiar24}) we compute (\ref{familiar23}) with the help of the following

\blem\label{lemamrein} Let $0<\alpha<\frac{\pi}{2}$ \z{and $\delta>0$ be given. Suppose
that $\phi:\RRR^3\to\CCC$ is a function with support in the sector
$P_{\be_3}^{\alpha}:=\{\bk\in\RRR^3:\bk\cdot\be_3>k\cos\alpha\}$ such that}
$\int\limits_{k=\delta}|\phi(\bk)|^2d\Omega(\bk)<\infty.$ Then
\begin{equation}\label{amrein}
  \int\limits_{\RRR^2}\left|\frac{1}{2\pi}\int\limits_{k=\delta}
e^{-i\bk\cdot\by}\phi(\bk)d\Omega(\bk)\right|^2d^2y=\int\limits_{k=\delta}|\phi(\bk)|^2\frac{1}{kk_3}d\Omega(\bk).
\end{equation}

\elem

\brem This lemma is proved in \cite{amrein:77}, Lemma 7.17.
The integration over the impact parameter is crucial for the
derivation and is a standard ingredient in the derivation of the
scattering cross section.\erem

Because of Corollary \ref{cort}, $T(\bk,\bkp)$ is bounded on
$\RRR^3\times\RRR^3$ and continuous on
$\RRR^3\times\RRR^3\setminus\{0\}$. Moreover,
$\widehat{\psieps}(\bk)\in\schwarz$ and
$\widehat{\psieps}(\bk)f_1(\bk)$ has support in
 $P_{\be_3}^{\vartheta_2}$ with $0<\vartheta_2<\frac{\pi}{2}$. Hence, by
Lemma \ref{lemamrein}, (\ref{familiar23}) becomes
\begin{align}\label{familiar3}
 \siglim(\Sigma)=&\lim\limits_{\epsilon\to 0}16\pi^4\int\limits_{C_\Sigma}
\int\limits_{k'=k}\left|T(\bk,\bkp)\right|^2\left|\widehat{\psieps}(\bkp)\right|^2\left|f_1(\bkp)\right|^2
  \frac{1}{\cos\vartheta'}d\Omega(\bkp)d^3k\notag\\=&\lim\limits_{\epsilon\to 0}16\pi^4\int\limits_{\Sigma}
\int\limits_{\RRR^3}\left|T(k'\bom,\bkp)\right|^2\left|\widehat{\psieps}(\bkp)\right|^2\left|f_1(\bkp)\right|^2
  \frac{1}{\cos\vartheta'}d^3k'd\Omega,
\end{align}
where $k'_3=k\cos\vartheta'.$ \z{Because $\supp f_1(\bk)\subset
P_{\be_3}^{\vartheta_2}$} with $0<\vartheta_2<\frac{\pi}{2},$ there exists a
$\delta>0$ such that $\delta<\cos\vartheta'.$ Hence the integral in
(\ref{familiar3}) is finite (it is $\leq \con\|\psi\|^2$). Thus, since
clearly $|\widehat{\psieps}(\bk)|^2\to\delta(\bk-\bko)$ (\z{in the sense that}
$\lim\limits_{\epsilon\to 0}\int |\widehat{\psieps}(\bk)|^2 g(\bk)d^3k=g(\bko)$
for any \z{bounded continuous function $g$),} and since $T(k'\bom,\bkp)$,
$f_1(\bkp)$ and $\frac{1}{\cos\vartheta'}$ are bounded and continuous as
functions of $\bkp$, we may conclude that
\begin{equation}\label{familiar4}
\siglim(\Sigma)=16\pi^4\int\limits_{\Sigma}
  |T(k_0\bom,\bko)|^2d\Omega.
\end{equation}

The proof of Theorem \ref{thedasisses} and Theorem \ref{thedasisseslight} will thus be complete once we
establish (\ref{familiar24}).  Changing variables, (\ref{familiar24})
follows from
\begin{align}\label{familiar25}
\lim\limits_{\epsilon\to 0} \int\limits_{\RRR^3} \int\limits_{y_p\geq\frac{D}{2}}\frac{1}{\epsilon^{2d}}
\left|\hspace{3pt}\int\limits_{k'=k}e^{-i(k_1'\frac{y_1}{\epsilon^d}+k_2'\frac{y_2}{\epsilon^d}-k_3'\Leps)}T(\bk,\bkp)
\widehat{\psieps}(\bkp)f_1(\bkp)k'd\Omega(\bkp)\right|^2dy_1dy_2d^3k=0.
\end{align}
(\ref{familiar25}) is the content of
\blem\label{lemamreinmore}Let $V\in(V)_5$, $\psi\in\schwarz$ and
suppose that $k_0>0.$ Let $l>2$, $d>2l-3$ and \z{let $M$ be} given
by (to simplify the notation we interchange $\bk$
and $\bkp$)
\begin{align}\label{amreinmore0}
M=M(y_1,y_2,\bkp,\epsilon):=\int\limits_{k=k'}e^{-i(k_1\frac{y_1}{\epsilon^d}+k_2\frac{y_2}{\epsilon^d}-k_3\Leps)}T(\bkp,\bk)
\widehat{\psieps}(\bk)f_1(\bk)kd\Omega(\bk).
\end{align}
Then \z{for any $D>0$}
\begin{align}\label{amreinmore}
\lim\limits_{\epsilon\to 0} \int\limits_{\RRR^3} \int\limits_{\yp\geq D}\frac{1}{\epsilon^{2d}}
\left|M\right|^2dy_1dy_2d^3k'=0.
\end{align}

\elem 
\bpro{} \z{We will establish the following
inequality (\ref{amreinmore2}) giving a bound on $M$:} There
exists a $c<\infty$ such that
\begin{align}\label{amreinmore2}
\left|M\right|^2\leq&
c\charac_{\left(\frac{k_0}{2},\frac{3}{2}k_0\right)}(k')\frac{\epsilon^{4d+5-4l}}{\yp^4}\frac{1}{\left(1+\frac{|k'-k_0|}{\epsilon}\right)^2}.
\end{align}
Assuming (\ref{amreinmore2}) we show now that (\ref{amreinmore}) 
follows. Using (\ref{amreinmore2}), \z{the integral in} (\ref{amreinmore}) is dominated by 
\begin{align}\label{amreinmore4}
\int\limits_{\frac{k_0}{2}< k'<\frac{3}{2}k_0}\int\limits_{\yp\geq
D}c\frac{\epsilon^{2d+5-4l}}{\yp^4}\frac{1}{\left(1+\frac{|k'-k_0|}{\epsilon}\right)^2}d^2yd^3k'\leq&\
c\epsilon^{2d+5-4l}\int\limits_{-\infty}^{\infty}\frac{dk'}{\left(1+\frac{|k'-k_0|}
{\epsilon}\right)^2}\notag\\=&\ c\epsilon^{2d+6-4l}\int\limits_{-\infty}^{\infty}\frac{dk'}{\left(1+|k'|\right)^2}
\notag\\=&\ c\epsilon^{2d+6-4l}.
\end{align}
\z{Since $d>2l-3$ there is a $\delta>0$ such that
$d=2l-3+\delta.$} Then (\ref{amreinmore4}) is of order
$\epsilon^{2\delta}$ and (\ref{amreinmore}) follows.

\newpage It thus remains to establish (\ref{amreinmore2}). Changing variables in
(\ref{amreinmore0}) from $\bom$ to $k_1,k_2$ we obtain\z{, with the
Jacobian determinant $k'k_3$ with
$k_3=k_3(k_1,k_2)=\sqrt{k'^2-k_1^2-k_2^2}$ and
$\bkplus=(k_1,k_2,k_3(k_1,k_2))$,}
\begin{align}\label{ammo2}
M=&\underset{k_1^2+k_2^2\leq k'^2}{\int\int}
e^{-i(k_1\frac{y_1}{\epsilon^d}+k_2\frac{y_2}{\epsilon^d}-k_3\Leps)}T(\bkp,\bkplus)
\widehat{\psieps}(\bkplus)f_1(\bkplus)k'\frac{1}{k'k_3}dk_1dk_2\notag\\=&\frac{1}{\epsilon^{\frac{3}{2}}}\underset{k_1^2+k_2^2\leq k'^2}{\int\int}
e^{-i(k_1\frac{y_1}{\epsilon^d}+k_2\frac{y_2}{\epsilon^d})}\left(T(\bkp,\bkplus)
\widehat{\psi}\left(\frac{\bkplus-\bko}{\epsilon}\right)e^{ik_3\Leps}\frac{f_1(\bkplus)}{k_3}\right)dk_1dk_2\notag\\=:&\frac{1}{\epsilon^{\frac{3}{2}}}\underset{k_1^2+k_2^2\leq k'^2}{\int\int}
e^{-i(k_1\frac{y_1}{\epsilon^d}+k_2\frac{y_2}{\epsilon^d})}g(k_1,k_2,\bkp,\epsilon)dk_1dk_2.
\end{align}
Performing two integration by parts \z{with respect} to $\bkpl:=(k_1,k_2)$, we obtain (using the fact that $f_1(\bkplus)$ and its derivatives vanish on the boundary of the region of integration) that
\begin{align}\label{ammo3}
|M|=&\frac{1}{\epsilon^{\frac{3}{2}}}\epsilon^d\left|\underset{k_p\leq k'}{\int\int}
\left(\nabla_{\bkpl}e^{-i(k_1\frac{y_1}{\epsilon^d}+k_2\frac{y_2}{\epsilon^d})}\right)\cdot\frac{\byp}{\yp^2}f_1(\bkplus)g(k_1,k_2,\bkp,\epsilon)dk_1dk_2\right|\notag\\=&\frac{1}{\epsilon^{\frac{3}{2}}}\epsilon^d\left|\underset{k_p\leq k'}{\int\int}
e^{-i(k_1\frac{y_1}{\epsilon^d}+k_2\frac{y_2}{\epsilon^d})}\frac{\byp}{\yp^2}\cdot\nabla_{\bkpl}g(k_1,k_2,\bkp,\epsilon)dk_1dk_2\right|\notag\\=&\frac{1}{\epsilon^{\frac{3}{2}}}\epsilon^{2d}\left|\underset{k_p\leq k'}{\int\int}\left(\nabla_{\bkpl}
e^{-i(k_1\frac{y_1}{\epsilon^d}+k_2\frac{y_2}{\epsilon^d})}\right)\cdot\frac{\byp}{\yp^2}\frac{\byp}{\yp^2}\cdot\nabla_{\bkpl}g(k_1,k_2,\bkp,\epsilon)dk_1dk_2\right|\notag\\=&\frac{1}{\epsilon^{\frac{3}{2}}}\epsilon^{2d}\left|\underset{k_p\leq k'}{\int\int}
e^{-i(k_1\frac{y_1}{\epsilon^d}+k_2\frac{y_2}{\epsilon^d})}\frac{\byp}{\yp^2}\cdot\nabla_{\bkpl}\frac{\byp}{\yp^2}\cdot\nabla_{\bkpl}g(k_1,k_2,\bkp,\epsilon)dk_1dk_2\right|\notag\\\leq&\frac{1}{\epsilon^{\frac{3}{2}}}\frac{\epsilon^{2d}}{\yp^2}\underset{k_p\leq k'}{\int\int}
\sum\limits_{i,j=1}^{2}\left|\partial_{k_i}\partial_{k_j}g(k_1,k_2,\bkp,\epsilon)\right|dk_1dk_2.
\end{align}

\z{We estimate now} the derivatives of $g$ on the support of $f_1$. \z{Note
  first that on $\supp f_1$ $k_3>k_0/2$. Using Corollary \ref{cort} we have
  for $i,j=1,2$ that}
\begin{align}\label{ammo8}
&\sup\limits_{\bkp\in\RRR^3,\bkplus\in\supp f_1}|T(\bkp,\bkplus)|\leq
\con,\;\sup\limits_{\bkp\in\RRR^3,\bkplus\in\supp f_1}\left|\partial_{k_i}T(\bkp,\bkplus)\right|\leq
\con,\\\notag&\sup\limits_{\bkp\in\RRR^3,\bkplus\in\supp
f_1}\left|\partial_{k_i}\partial_{k_j}T(\bkp,\bkplus)\right|\leq \con.
\end{align}
To estimate the wave function
$\widehat{\psi}\left(\frac{\bkplus-\bko}{\epsilon}\right)$ \z{and its
derivatives} we introduce the following notation:
\begin{align}\label{ammo8a}
P_k:=\frac{1}{1+\frac{|k-k_0|}{\epsilon}},\;P_{\bk}:=\frac{1}{1+\frac{|\bk-\bko|}{\epsilon}}.
\end{align}
Clearly
\begin{align}\label{ammo8b}
P_{\bk}\leq P_k.
\end{align}
Since $\psi\in\schwarz$, $\widehat{\psi}$ and its derivatives decay faster
than \z{the reciprocal of any polynomial, we can find for $\bkplus\in\supp
f_1$ and for $n\in\NNN$ suitable constants such that}
\begin{align}\label{ammo9}
&\left|\widehat{\psi}\left(\frac{\bkplus-\bko}{\epsilon}\right)\right|\leq
\con P_{\bkplus}^n,\;\left|\partial_{k_i}\widehat{\psi}\left(\frac{\bkplus-\bko}{\epsilon}\right)\right|\leq\frac{\con}{\epsilon}
P_{\bkplus}^n,\;\left|\partial_{k_i}\partial_{k_j}\widehat{\psi}\left(\frac{\bkplus-\bko}{\epsilon}\right)\right|
\leq\frac{\con}{\epsilon^2}P_{\bkplus}^n.
\end{align}
The \z{derivatives of the} third factor $e^{-ik_3\Leps}$ of $g$ can be estimated on $\supp f_1$ as follows:
\begin{align}\label{ammo11}
\left|e^{-ik_3\Leps}\right|\leq 1,\;\left|\partial_{k_i}e^{-ik_3\Leps}\right|\leq\Leps\frac{|k_i|}{|k_3|}\leq\Leps |k_i|.
\end{align}
\z{Since $|k_i|P_{\bkplus}\leq \epsilon$, we obtain using (\ref{ammo9}) with $n=j+1$ and (\ref{leps})  that} 
\begin{align}\label{ammo11a}
\left|\left(\partial_{k_i}e^{-ik_3\Leps}\right)\widehat{\psi}\left(\frac{\bkplus-\bko}{\epsilon}\right)\right|\leq \con\Leps |k_i| P_{\bkplus}P_{\bkplus}^j\leq \con\Leps \epsilon P_{\bkplus}^j=\frac{\con}{\epsilon^{l-1}}P_{\bkplus}^j,\;j\text{ arbitrary}.
\end{align}
With a similar calculation we find that
\begin{align}\label{ammo12}
\left|\left(\partial_{k_i}\partial_{k_j}e^{-ik_3\Leps}\right)\widehat{\psi}\left(\frac{\bkplus-\bko}{\epsilon}\right)\right|
\leq\frac{\con}{\epsilon^{2l-2}}P_{\bkplus}^j,\;j\text{ arbitrary},
\end{align}
and analogous estimates for terms which contains derivatives of $\widehat{\psi}\left(\frac{\bkplus-\bko}{\epsilon}\right).$
\z{Clearly we have that}
\begin{align}\label{ammo13}
\sup\limits_{\bkplus\in\supp f_1}\left|\frac{f_1(\bkplus)}{k_3}\right|\leq \con,\;\sup\limits_{\bkplus\in\supp f_1}\left|\partial_{k_i}\frac{f_1(\bkplus)}{k_3}\right|\leq \con,\;\sup\limits_{\bkplus\in\supp f_1}\left|\partial_{k_i}\partial_{k_j}\frac{f_1(\bkplus)}{k_3}\right|\leq \con,\;i,j=1,2.
\end{align}
Combining (\ref{ammo8}), (\ref{ammo9})-(\ref{ammo13}) and using that
$2l-2>2$ since $l>2$ we obtain for \z{all $\bkp\in\RRR^3$ and any
  $n\in\NNN$ that 
\begin{align}\label{ammo14}
\left|\partial_{k_i}\partial_{k_j}g(k_1,k_2,\bkp,\epsilon)\right|\leq\frac{\con}{\epsilon^{2l-2}}P_{\bkplus}^n,
\end{align}
for all $(k_1,k_2)$ such that $\bkplus\in\supp f_1$.}

Reintroducing the original \z{integration variable $\bom$  we} then have that
\begin{align}\label{ammo4}
|M|\leq&\frac{\con}{\yp^2}\epsilon^{2d-2l+\frac{1}{2}}\int\limits_{k=k'}\chi_{\{f_1>0\}}P_{\bk}^nk'k_3d\Omega(\bk)\notag\\ \leq&\frac{\con}{\yp^2}\epsilon^{2d-2l+\frac{1}{2}}\charac_{\left(\frac{k_0}{2},\frac{3}{2}k_0\right)}(k')\int\limits_{k=k',|\bk-\bko|<\frac{k_0}{2}}P_{\bk}^nd\Omega(\bk).
\end{align}
Choosing $n=4$ in (\ref{ammo4}) and splitting $P_{\bk}^4$ into
\begin{align}\label{ammo22}
P_{\bk}^4=P_{\bk}^1P_{\bk}^3\leq P_{k}^1P_{\bk}^3
\end{align}
we obtain \z{that}
\begin{align}\label{ammo14aa}
|M|\leq&\frac{\con}{\yp^2}\epsilon^{2d+\frac{1}{2}-2l}\charac_{\left(\frac{k_0}{2},\frac{3}{2}k_0\right)}(k')P_{k'}^1\int\limits_{k=k',|\bk-\bko|<\frac{k_0}{2}}P_{\bk}^3d\Omega(\bk).
\end{align}
Moreover, it is easy to see that
\begin{align}\label{ammo14aaa}
\int\limits_{k=k',|\bk-\bko|<\frac{k_0}{2}}P_{\bk}^3d\Omega(\bk)
\leq \con\int\limits_{\RRR^2}\frac{1}{\left(1+\frac{k_p}{\epsilon}\right)^3}dk_1dk_2\leq
\con\epsilon^2.
\end{align}
Thus
\begin{align}\label{ammo14aaaa}
|M|\leq&\frac{\con}{\yp^2}\epsilon^{2d+\frac{5}{2}-2l}\charac_{\left(\frac{k_0}{2},\frac{3}{2}k_0\right)}(k')P_{k'}^1
\end{align}
and (\ref{amreinmore2}) follows. This completes the proof of Lemma
\ref{lemamreinmore}. \epro

\section{Summary and \z{outlook}}\label{chproblem}

The purpose of this paper has been to rigorously derive the standard
formula for the scattering cross section starting from a microscopic
model of a scattering experiment. While the use of Bohmian mechanics
is crucial for our result, we would like to stress that major parts
of our proof are vital even from an orthodox point of view. These
parts concern in particular the replacement of the incoming asymptote by its
scattering state (cf. Lemma \ref{lemint} and Remark \ref{remint}) and the
flux-across-surfaces theorem in a formulation which depends only on
the smoothness of the scattering state (cf. Proposition
\ref{propflux}, Lemma \ref{lemmapping} and \cite{duerr1:04}).

Several problems have been left for future work, which we shall
mention here.

\begin{itemize}
\item Bound states: Our assumption A\ref{condbound} arises from
the problem that in general the translation of the initial wave
function by the impact parameter $\by$---which is needed for the
averaging over the beam profile---will produce wave functions
which have a component in the bound states. One would then have to
show that asymptotically the crossing statistics are induced by
the ``relevant part'' $\psi'$ of the wave function, namely
\begin{equation*}
  \psi':=P\psi,
\end{equation*}
where $P$ is the projection onto the absolutely continuous
subspace $\mathcal{H}_{\text{a.c.}}(H)$ and is given by
\begin{equation*}
  P:=\Omega_-\Omega_-^*.
\end{equation*}

Note that by using Lemma \ref{lemint} one can also show that
\begin{equation}\label{ppsi1}
\lim\limits_{L\to\infty}\int\limits_{\GammaL}\|P\psi_{\by}-\psi_{\by}\|^2d^2y=0,
\end{equation}
i.e., that the bound state component is small in an $L^2$-sense.
This is however not directly applicable.
\item It would of course
be desirable to derive the crossing statistics for many particles
guided in general by an entangled wave function both for the
noninteracting case and eventually even for interacting particles
\cite{duerrteufel:02}.

\item \z{We are currently working \cite{duerr4:04} on a detailed
  formulation of the conditions characterizing the scattering regime, which
  turns out to be surprisingly intricate. What we have shown here is that
  the simplest limiting procedure that brings the experimental arrangement
  into the scattering regime yields the standard formula of formal
  scattering theory. This formula should of course hold much more
  generally---more or less for all limits corresponding to the scattering
  regime---but establishing that this is so remains a formidable challenge.}

\end{itemize}

\section{Appendix}

\bpro{ of Lemma \ref{lemmapping}} Let $\psi\in\g.$ Then
there is a $\chi\in\gprime$ and a $t\in\RRR$
such that
\begin{equation*}
\psi=e^{-iHt}\chi.
\end{equation*}
Using the intertwining property
(\ref{inter}) we obtain
\begin{equation}\label{map0}
\pout=\Omega_+^{-1}\psi=\Omega_+^{-1}e^{-iHt}\chi=e^{-iH_0t}\Omega_+^{-1}\chi=e^{-iH_0t}\cout.
\end{equation}
Since $\ghut$ is invariant under time shifts it
suffices to show that $\couth(\bk)$ is in $\ghut.$
 Since $\lxr^2H^n\chi(\bx)\in
L_2(\RRR^3),\;0\leq n\leq 8,$ and $\lxr^4H^n\chi(\bx)\in
L_2(\RRR^3),\;0\leq n\leq 3,$ we have
\begin{align}\label{l1l2}
\begin{split}
H^n\chi(\bx)&\in L_1(\RRR^3)\cap L_2(\RRR^3),\;0\leq n\leq 8,\\
\lxr^jH^n\chi(\bx)&\in L_1(\RRR^3)\cap L_2(\RRR^3),\;0\leq n\leq
3,\;j=\{1,2\}.
\end{split}
\end{align}
Using Proposition \ref{propexpansion} (ii), (iii) we have for
$f\in\ltr$:
\begin{align}\label{mapa}
\mathcal{F}_+\Omega_+f=&\mathcal{F}f,
\end{align}
and hence for $\chi=\Omega_+\cout$ we have that
\begin{align}\label{mapaa}
\couth(\bk)=&\mathcal{F}_+\chi(k)=(2\pi)^{-\frac{3}{2}}\int\varphi^*_+(\bx,\bk)\chi(\bx)d^3x.
\end{align}
Using the intertwining property (\ref{inter}) we thus have:
\begin{align}\label{map}
\frac{k^2}{2}\couth(\bk)&=\widehat{H_0\chi}_{\text{out}}(\bk)=\mathcal{F}(H_0\Omega_+^{-1}\chi)(\bk)=\mathcal{F}(\Omega_+^{-1}H\chi)(\bk)=\mathcal{F}_+(H\chi)(\bk)\notag\\&=(2\pi)^{-\frac{3}{2}}\int\varphi^*_+(\bx,\bk)(H\chi)(\bx)d^3x.
\end{align}

Similarly, applying $H_0^n$ to $\couth(\bk)$ ($0\leq n\leq 8$) we
obtain
\begin{align}\label{map1}
\frac{k^{2n}}{2^n}\couth(\bk)&=(2\pi)^{-\frac{3}{2}}\int\varphi^*_+(\bx,\bk)(H^n\chi)(\bx)d^3x.
\end{align}
Since the generalized eigenfunctions are bounded (Proposition
\ref{proplippschw} (ii)) and $H^n\chi\in L_1(\RRR^3),\;$ $0\leq
n\leq 8,$ we obtain
\begin{equation}\label{map2}
|\couth(\bk)|\leq \con(1+k)^{-16}\leq \con(1+k)^{-15}.
\end{equation}
Because of Proposition \ref{proplippschw} (iii) and (\ref{l1l2})
we can differentiate (\ref{mapaa}) \z{with respect to }  $k_i$ and get
\begin{equation}\label{map5a}
\left|\partial_{k_i}\couth(\bk)\right|=\left|(2\pi)^{-\frac{3}{2}}\int\left(\partial_{k_i}\varphi^*_+(\bx,\bk)\right)\chi(\bx)d^3x\right|\leq
\con,\;\forall\bk\in\RRR^3\setminus\{0\}.
\end{equation}
Differentiating (\ref{map1}) with $n=3$ \z{with respect to }  $k_i$ we obtain
\begin{equation}\label{map5}
k^6\partial_{k_i}\couth(\bk)=8(2\pi)^{-\frac{3}{2}}\int\left(\partial_{k_i}\varphi^*_+(\bx,\bk)\right)(H^3\chi)(\bx)d^3x-6k^5\couth(\bk)\frac{k_i}{k}.
\end{equation}
Again the right-hand side is bounded because of Lemma
\ref{proplippschw} (iii), (\ref{l1l2}) and (\ref{map2}). Hence, we
obtain with (\ref{map5a}):
\begin{equation}\label{map6}
\left|\partial_{k_i}\couth(\bk)\right|\leq
\con(1+k)^{-6},\;\forall\bk\in\RRR^3\setminus\{0\}.
\end{equation}
Using Proposition \ref{proplippschw} (iii) and (\ref{mapaa}) we
may control $\kfac$ times a second derivative of $\couth(\bk)$,
obtaining
\begin{equation}\label{map6a}
\left|\kfac\partial_{k_j}\partial_{k_i}\couth(\bk)\right|=\left|(2\pi)^{-\frac{3}{2}}\int\left(\kfac\partial_{k_j}\partial_{k_i}\varphi^*_+(\bx,\bk)\right)\chi(\bx)d^3x\right|\leq
\con,\;\forall\bk\in\RRR^3\setminus\{0\}.
\end{equation}
For the last inequality we have also used (\ref{l1l2}) with $j=2$
and $n=0$. Similarly, using (\ref{map5}) we obtain
\begin{align}\label{map6b}
k^6\kfac\partial_{k_j}\partial_{k_i}\couth(\bk)=&8(2\pi)^{-\frac{3}{2}}\int\left(\kfac\partial_{k_j}\partial_{k_i}\varphi^*_+(\bx,\bk)\right)(H^3\chi)(\bx)d^3x\notag\\&-30k^4\frac{k_j}{k}\frac{k_i}{k}\kfac\couth(\bk)-6k^5\frac{k_i}{k}\kfac\partial_{k_j}\couth(\bk)\notag\\&-6k^5\couth(\bk)\kfac\frac{k\delta_{ij}k-k_ik_j}{k^3}-6k^5\frac{k_j}{k}\kfac\partial_{k_i}\couth(\bk),
\end{align}
with right-hand side that is bounded because of Proposition
\ref{proplippschw} (iii), (\ref{l1l2}), (\ref{map2}) and
(\ref{map6}). Hence, using (\ref{map6a}),
\begin{equation}\label{map7}
\left|\kfac\partial_{\bk}^{\alpha}\couth(\bk)\right|\leq
\con(1+k)^{-6}\leq
\con(1+k)^{-5},\;|\alpha|=2,\;\forall\bk\in\RRR^3\setminus\{0\}.
\end{equation}
Equation (\ref{map6}) implies also that
\begin{equation}\label{mapb}
\left|\partial_{k}\couth(\bk)\right|\leq
\con(1+k)^{-6},\;\forall\bk\in\RRR^3\setminus\{0\}.
\end{equation}
Similarly, twice differentiating (\ref{mapaa}) \z{with respect to }  $k$ we
obtain that
\begin{equation}\label{mapbzu}
\left|\partial_{k}^2\couth(\bk)\right|\leq
\con,\;\forall\bk\in\RRR^3\setminus\{0\},
\end{equation}
and then twice differentiating
(\ref{map1}) for $n=2$ \z{with respect to }  $k$ we obtain
\begin{equation}\label{mapc}
\left|\partial_{k}^2\couth(\bk)\right|\leq \con(1+k)^{-4}\leq
\con(1+k)^{-3},\;\forall\bk\in\RRR^3\setminus\{0\},
\end{equation}
using Proposition \ref{proplippschw} (iv), (\ref{l1l2}),
(\ref{map2}), (\ref{mapb}) and (\ref{mapbzu}).

With (\ref{map2}), (\ref{map6}), (\ref{map7}) and (\ref{mapc}) we see that $\couth(\bk)\in\ghut.$ \epro

\bpro{ of Lemma \ref{lemuniform}} In the proof of Proposition
\ref{propflux} in \cite{duerr1:04} the absolute value of the flux
integrated over time and the surface $ RS^2$ with $R>R_0$
(with some $R_0>0$ depending on the potential) is shown to be
bounded (uniformly in $R$) by linear combinations of integrals involving
$\widehat{\psi}_{\text{out}}(\bk)$ and its derivatives, namely
integrals over expressions corresponding to the left hand side of
the inequalities in Definition \ref{defg}. Thus these bounds are
finite if $\pouth(\bk)\in\ghut.$ To bound the integrated flux uniformly for all $\psieps_{\by},\;\by\in\Aeps$ (and $\epsilon$ small enough and fixed), $\mathcal{F}\left(\psieps_{\by,\text{out}}\right)(\bk)=\mathcal{F}\left(\Omega_+^{-1}\psieps_{\by}\right)(\bk)$
(note that $\psieps_{\by}\in\mathcal{H}_{\text{a.c.}}(H)$, for all
$\by\in\Aeps$, cf. (i) in Definition \ref{defpotstrong} or \ref{defpotstronglight}) must be bounded as in Definition \ref{defg} with
constants uniform in $\by\in\Aeps.$ These constants depend, according
to the proof of Lemma \ref{lemmapping}, on the norms of
\begin{equation}\label{mapnorm}
\|H^n\psieps_{\by}\|_1,\;0\leq n\leq 8\text{ and
}\|\lxr^jH^n\psieps_{\by}\|_1,\;0\leq n\leq 3,\;j\in\{1,2\}.
\end{equation}
We will show that for $\epsilon$ small enough there exists a constant $C>0$ such that
\begin{equation}\label{major}
|H^n\psieps_{\by}(\bx)|\leq C(1+x)^{-6},\;0\leq n\leq 8,\;\forall\by\in\Aeps.
\end{equation}
Thus the norms in (\ref{mapnorm}) are bounded uniformly in $\by\in\Aeps$ and Lemma \ref{lemuniform} follows.

It remains to establish (\ref{major}). We start with $n=0.$ Since
$\psi\in\schwarz$ and $\by\in\Aeps,\;\Aeps$ compact,  we obtain
\begin{equation}\label{mapnorma}
|\psieps_{\by}(\bx)|=\epsilon^{\frac{3}{2}}|\psi(\epsilon(\bx-\by))|\leq\con(1+|\bx-\by|)^{-6}\leq
\con(1+x)^{-6},\;\forall\by\in\Aeps.
\end{equation}
For $n=1$ we have with $\psieps_{\by}\equiv T_{\by}\psieps$ ($T_{\by}$
is the translation operator) and $[T_{\by},H_0]_-=0$ 
\begin{align}\label{mapnormb}
|H\psieps_{\by}(\bx)|=&|(H_0+V)T_{\by}\psieps(\bx)|=|T_{\by}H_0\psieps(\bx)|+\epsilon^{\frac{3}{2}}|V(\bx)\psi(\epsilon(\bx-\by))|.
\end{align}
Using now $|V(\bx)|<M<\infty$ for $V\in\mathcal{V}$ or $\sup\limits_{\bx\in\supp \psieps_{\by}}|V(\bx)|<M<\infty$ for $\psi\in C_0^\infty(\RRR^3)$, $V\in\mathcal{V'}$, $\by\in\Aeps$ and $\epsilon$ small enough, we obtain together with (\ref{mapnorma})
\begin{align}\label{mapnormbb}
|H\psieps_{\by}(\bx)|\leq |T_{\by}H_0\psieps(\bx)|+\con(1+x)^{-6}.
\end{align}
Since $\psieps\in\schwarz$ we have that also $H_0\psieps\in\schwarz$ so
that analogously to (\ref{mapnorma}), there is the bound
\begin{align}\label{mapnormc}
|T_{\by}H_0\psieps(\bx)|\leq \con(1+x)^{-6},\;\forall\by\in\Aeps.
\end{align}
Equations (\ref{mapnormbb}) and (\ref{mapnormc}) yield (\ref{major}) for
$n=1.$ Analogously, we obtain (\ref{major}) for $2\leq n\leq 8$ by
using the fact that $\psi\in\schwarz$ and $|\partial_{\bx}^{\alpha}V(\bx)|<M<\infty,\;\forall\;|\alpha|\leq
14$, if $V\in\mathcal{V}$ or $\sup\limits_{\bx\in\supp \psieps_{\by}}|\partial_{\bx}^{\alpha}V(\bx)|<M<\infty,\;\forall\;|\alpha|\leq
14$, for all $\by\in\Aeps$ and $\epsilon$ small enough if $\psi\in C_0^\infty(\RRR^3)$ and $V\in\mathcal{V'}$.\epro

\bigskip

\noindent \z{\textit{Acknowledgments.} The work of S.~Goldstein was
supported in part by NSF Grant DMS-0504504. The work of T. Moser was
supported by the DFG (DU 120/10).} The work of N.~Zangh\`\i\ was supported by INFN.

\end{document}